\documentclass[10pt,twocolumn,showpacs,prb,aps,floatfix]{revtex4-1}

\usepackage{graphicx}
\usepackage{dcolumn}
\usepackage{float}
\usepackage{amsmath}
\usepackage{amssymb}
\usepackage{bm}

\newcommand{\comment}[1]{}
\renewcommand{\epsilon}{\varepsilon}
\DeclareMathOperator{\sgn}{sgn}

\begin{document}


\title{Magneto-optics of general pseudospin-$s$ two-dimensional Dirac-Weyl fermions}

\author{J.D. Malcolm}
\author{E.J. Nicol}
\affiliation{Department of Physics, University of Guelph,
Guelph, Ontario N1G 2W1 Canada} 
\affiliation{Guelph-Waterloo Physics Institute, University of Guelph, Guelph, Ontario N1G 2W1 Canada}
\date{\today}

\begin{abstract}
{The popularity of graphene--a pseudospin-$\frac{1}{2}$ two-dimensional Dirac-Weyl material--has prompted the search for related materials and the characterization of their properties.  In this work, the magneto-optical conductivity is calculated for systems that obey the general pseudospin-$s$ two-dimensional Dirac-Weyl Hamiltonian, with particular focus on $s=\left\{\frac{1}{2},1,\frac{3}{2},2\right\}$.  This generalizes calculations that have been made for $s=\frac{1}{2}$ and follows previous work on the optical response of these systems in zero field.  In the presence of a magnetic field, Landau levels condense out of the $2s+1$ energy bands.  As the chemical potential in a system is shifted, patterns arise in the appearance and disappearance of certain peaks within the optical spectra.  These patterns are markedly different for each case considered, creating unique signatures in the magneto-optics.  The general structure of each spectrum and how they compare is discussed.}
\end{abstract}


\maketitle

\section{Introduction}

Graphene, isolated in 2004,\cite{Novoselov04} is a two-dimensional (2D) crystal honeycomb lattice of carbon atoms.\cite{Zhang05}  At low energies, the Hamiltonian for the fermionic charge carriers in graphene maps onto the massless Dirac Hamiltonian with an effective speed of light\cite{Novoselov05} of $v_{F}\sim 10^{6}$ m/s.  This gives rise to the famous energy dispersion exhibiting a Dirac cone at each of the two distinct K points in the hexagonal Brillouin zone.  The Hamiltonian can be decomposed into the Weyl basis, whereby left- and right-handed Weyl 2-spinors will describe particles at the respective K points.  The reduced Hamiltonian, acting on only one of the Weyl spinors, is as follows, with dimension $\mathrm{dim}\,\hat{\mathcal{H}} = \mathbb{C}^{2s+1}\otimes\mathbb{R}^{2}$,
\begin{equation}\label{eqn:WHam}
 \hat{\mathcal{H}} = \hbar\,v_{c}\boldsymbol{S}\cdot\boldsymbol{k}.
\end{equation}
In Eq.~(\ref{eqn:WHam}), $v_{c}$ is a characteristic velocity such that the graphene Fermi velocity is $v_{F}=\frac{1}{2}v_{c}$ and $\boldsymbol{S}=\left( S_{x},S_{y} \right)$ are the first two spin-$\frac{1}{2}$ matrices.  In this way, graphene is referred to as an $s=\frac{1}{2}$ pseudospin Dirac-Weyl material (DW).  The lower and upper Dirac cones in the energy dispersion are labelled with the $S_{z}$ diagonal elements $\lambda=\left\{-\frac{1}{2},\frac{1}{2}\right\}$, respectively.  The pseudospin is not related to a magnetic moment as in the intrinsic spin of an electron.  However, unlike other spin analogues like the nucleon isospin, pseudospin is associated with an angular momentum.\cite{Mecklenburg11}

\begin{figure}[b]
 \begin{center}
  \includegraphics[width=1.0\linewidth]{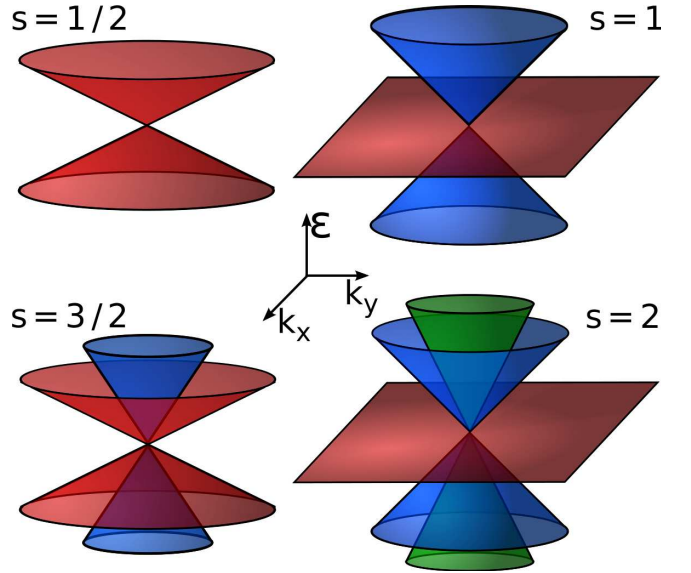}
 \end{center}
 \caption{\label{fig:cones}(Color online) Energy dispersion at a single K point in Dirac-Weyl materials with different values of pseudospin-$s$.}
\end{figure}

There are many exciting potential electronic applications for graphene, such as in flexible touch-screen displays\cite{Radivojevic12} and in solar cells.\cite{Miao12}  These particular examples illustrate the importance of understanding the material's optical properties.  Furthermore, scientists have been seeking out other 2D materials that may also lead to promising technologies.  The focus of this paper is on systems with a Hamiltonian of the form in Eq.~(\ref{eqn:WHam}), but now with general pseudospin-$s$, a class referred to as 2D Weyl materials.  The energy dispersion for a collection of Weyl materials is presented in Fig.~\ref{fig:cones} and is described mathematically by
\begin{equation}\label{eqn:eps_zero}
 \epsilon_{\lambda} = \lambda\hbar\,v_{c}\left|\boldsymbol{k}\right|.
\end{equation}
We see that the pseudospin-$s$ dispersion has $2s+1$ bands, in the form of nested Dirac cones, labelled by $\lambda=\left\{-s,-s+1,...,s\right\}$.  The Fermi velocity in a particular band is $v_{F} = \lambda v_{c}$.  Notably, materials for which $s$ is a whole number contain a zero-energy flat band.

In this work, we only go so far as to consider materials described by Eq.~(\ref{eqn:WHam}) and make no distinction as to the exact nature of the lattice giving rise to said Hamiltonian.  Candidates for Weyl materials have been theorized as a layering of $2s+1$ triangular lattices\cite{Dora11} and as an optical lattice of ultra cold $^{40}$K atoms.\cite{Lan11,Mazza12}

With the incorporation of graphene into optical devices, it is not a stretch to suppose that DW's with a different value for $s$ could have similar useful applications.  Characterizing the optical properties of a general-$s$ DW is crucial in determining the viability of future DW devices.  The calculation of the optical conductivity, $\sigma_{\alpha\beta}$, assumes that negligible momentum is provided to the system via excitation by a photon.  Thus, transitions drawn in k-space are drawn vertically, where there is no change in momentum.  As such, we can treat the different K-points (however many there may be in a particular DW) as isolated.  Thus, one can compute $\sigma_{\alpha\beta}$ of a pseudospin-$s$ DW by considering only the Weyl Hamiltonian, Eq.~(\ref{eqn:WHam}).  The only additional step is to include a degeneracy factor $g$ that counts the number of distinct K points as well as any other degeneracies.  For example, $g=4$ in graphene due to the two-fold spin and two-fold valley (K-point) degeneracies.  From this point on, we will permit our study to refer to the properties of DW materials, while only working with the simpler, yet sufficient Weyl Hamiltonian.

In the following, we calculate the magneto-optics of general-$s$ DW materials.  Previous work for the zero-field conductivity has been done by D\'{o}ra {\em et al}.\cite{Dora11}  The magneto-optics in graphene\cite{Gusynin07PRL,Gusynin07JP} ($s=\frac{1}{2}$)---and other lower dimensional systems\cite{Rose13,Tabert13PRL,Tabert13PRB,Li13}---have been calculated in other works.  The spectra ${\rm Re}\,\sigma_{xx}$, ${\rm Re}\,\sigma_{+}$, and ${\rm Re}\,\sigma_{-}$ can be measured directly in experiment, for instance as absorbance\cite{Mak12} via the relation $A=\frac{4\pi}{c}\sigma$, where $A$ is the absorbance and $c$ is the actual speed of light.  Such experiments have been performed on graphene in zero-field\cite{Mak08,Nair08,Li08,Wang08} and in the presence of an external magnetic field,\cite{Sadowski06,Sadowski07,Jiang07,Henriksen10} agreeing with the theory.

\section{Theory}
\label{sec:theory}

\begin{figure}
\includegraphics[width=1.0\linewidth]{bare.eps}
\caption{\label{fig:bare}(Color online) Landau level energies in (a)$s=2$ and (b)$s=\frac{3}{2}$ 2D Dirac-Weyl materials.  Each Landau level is characterized by a Fock number $n$ and represented by a red dot.  The dotted lines trace the energy bands, or more accurately energy ``branches,'' and are labelled by $S_{z}$ projections, $\lambda$.}
\end{figure}

We extend the calculation of the optical conductivity tensor, $\sigma_{\alpha\beta}\left(\Omega\right)$, as a function of photon energy, to include a nonzero magnetic field perpendicular to the plane of the material, $\boldsymbol{B}=\left(0,0,B\right)$.  Once the field is imposed, the kinetic momentum $\hbar\boldsymbol{k}$ is no longer conjugate to position.  A Peierls substitution is made $\hbar\boldsymbol{k} \rightarrow \hbar\boldsymbol{\Pi} = \hbar\boldsymbol{k} + \frac{e}{c}\boldsymbol{A}$ where $\hbar\boldsymbol{k}$ now represents the canonical momentum and is itself conjugate to position.  $\boldsymbol{A}$ is the vector potential such that $\boldsymbol{B} = \boldsymbol{\nabla}\times\boldsymbol{A}$ and we have assigned a charge of $-e$ to the particle-like fermions, as in graphene.  Under the substitution, one finds the commutation relation $\left[\Pi_{x},\Pi_{y}\right] = -\frac{i}{l_{B}^{2}}$, where $l_{B} = \sqrt{\frac{\hbar c}{e|B|}}$ is the magnetic length scale.  This prompts the definition of the operators
\begin{equation}\label{eqn:Fock}
 a^{\dagger} = \frac{1}{\sqrt{2}}l_{B}\left(\Pi_{x} + i\Pi_{y}\right)\; ,\; a = \left(a^{\dagger}\right)^{\dagger}.
\end{equation}
The resulting bracket $\left[a,a^{\dagger}\right]=1$ shows that $a^{\dagger}$ and $a$ are bosonic creation and annihilation operators acting on a Fock space $\mathcal{F}_{+} = \mathrm{span}\left\{\left|n\right\rangle,n=0,1,2,...\right\}$ (the $+$ subscript refers to the bosonic Fock space).  With these new operators, the Hamiltonian in Eq.~(\ref{eqn:WHam}) can be rewritten\cite{Lan11} as a tridiagonal matrix using factors $\rho_{m} = \sqrt{m\left(2s+1-m\right)}$ as
\begin{equation}\label{eqn:WHamB}
\begin{split}
\hat{\mathcal{H}}& = \gamma \left( S_{+}a + S_{-}a^{\dagger} \right)\\
 & = \gamma 
     \begin{pmatrix}
      0 & \rho_{1}a & & & \\
      \rho_{1}a^{\dagger} & 0 & \rho_{2}a & &  \\
       & \rho_{2}a^{\dagger} & \ddots & \ddots &  \\
       & & \ddots & 0 & \rho_{2s}a \\
       & & & \rho_{2s}a^{\dagger} & 0 \\
     \end{pmatrix}
\end{split}
\end{equation}
$S_{\pm} = S_{x} \pm i S_{y}$ are the pseudospin ladder operators and $\gamma = \frac{\hbar v_{c}}{\sqrt{2}l_{B}}$ is a measure of energy.  The dimension of Eq.~(\ref{eqn:WHamB}) is $\mathrm{dim}\:\hat{\mathcal{H}} = \mathbb{C}^{2s+1}\otimes\mathcal{F}_{+}\otimes\mathcal{F}_{+}$.  Note that the $\mathbb{R}^{2}$ subspaces associated with the $\boldsymbol{k}$ components in Eq.~(\ref{eqn:WHam}) have now each been quantized into $\mathcal{F}_{+}$.\cite{Goerbig11}

In the presence of a magnetic field, the continuous bands seen in Fig.~\ref{fig:cones} become quantized into discrete Landau levels.\cite{Zhang05,Lan11}  The projection of a state in one of the Fock spaces remains a constant of motion, leaving no bearing on the energy.\cite{Goerbig11}  This makes the Landau levels highly degenerate.  These energy states can be grouped into a collection of $2s+1$ ``bands," indexed by $\lambda$ (retaining the $S_{\rm z}$ labels).  Note that in this regime, $\lambda$ is no longer a good energy quantum number (found in Eq.~(\ref{eqn:eps_zero})), and is instead only used here to help categorize the different states.  Within a particular band, each Landau level is indexed by a Fock number, $n$ (see Fig.~\ref{fig:bare}).  The general form of the Landau level wavefunction is given below and the energies are given in Table  {\ref{tab:Landau} for pseudospin values up to $s=2$ (reproduced from work by Lan \textit{et al}.\cite{Lan11}),
\begin{equation}\label{eqn:ket}
 \left|\lambda,n\right\rangle = \left( \alpha^{1}_{\lambda n}\left|n-2s\right\rangle,\alpha^{2}_{\lambda n}\left|n-2s+1\right\rangle,...,\alpha^{2s+1}_{\lambda n}\left|n\right\rangle\right)^{\rm T}.
\end{equation}
$\alpha^{i}_{\lambda n}$ is a c-number, the form of which has been derived recursively and is defined in Appendix \ref{sec:Alphas}.  Note that the superscript $i$ indexes the component of the $\mathbb{C}^{2s+1}$ spinor space; it is not an exponent.

\renewcommand{\arraystretch}{1.5}
\begin{table}
\caption{\label{tab:Landau} Energies for the Landau levels found in a pseudospin-$s$ 2D Dirac-Weyl material in the presence of a magnetic field, in units $\gamma=\frac{\hbar v_{c}}{\sqrt{2}l_{B}}$.  Each level is labelled by a band index, $\lambda$, and Fock number, $n$.  Within each band certain low values of $n$ may be excluded from the set, as indicated in the final column.  This table is a reproduction from the work by Lan \textit{et al}.\cite{Lan11}}
\centering
\begin{tabular}{c l l}
\hline\hline
$s$ & $\epsilon_{\lambda n}$ & Allowed $n$ \\
\hline
$\frac{1}{2}$ & $\epsilon_{\pm \frac{1}{2},n} = \pm\gamma\sqrt{n}$ & $n=0,1,...$ \\
$1$ & $\epsilon_{0,n} = 0$ & $n=0,2,3,...$ \\
 & $\epsilon_{\pm 1,n} = \pm\gamma\sqrt{2\left(2n-1\right)}$ & $n=1,2,...$ \\
$\frac{3}{2}$ & $\epsilon_{\pm\frac{1}{2},n} = \pm\gamma\sqrt{5\left(n-1\right)-\sqrt{16\left(n-1\right)^{2}+9}}$ & $n=2,3,...$ \\
 & $\epsilon_{\pm\frac{3}{2},n} = \pm\gamma\sqrt{5\left(n-1\right)+\sqrt{16\left(n-1\right)^{2}+9}}$ & $n=0,1,...$ \\
$2$ & $\epsilon_{0,n} = 0$ & $n=0,2,4,5,...$ \\
 & $\epsilon_{\pm 1,n} = \pm\gamma\sqrt{5(2n-3)-3\sqrt{4n^{2}-12n+17}}$ & $n=3,4,...$ \\
 & $\epsilon_{\pm 2,n} = \pm\gamma\sqrt{5(2n-3)+3\sqrt{4n^{2}-12n+17}}$ & $n=1,2,...$ \\
\hline\hline
\end{tabular}
\end{table}

As seen in the final column of Table \ref{tab:Landau}, some low values of $n$ do not exist in certain bands.  This is explained in more detail in the appendix.  However, referring to Eq.~(\ref{eqn:ket}), it can at least be seen why the number of levels at a particular $n$ must be limited.  First, consider a state with $n=0$.  Because $\left|n\right\rangle$ is only defined for non-negative values of $n$, the wavefunction is simply
\begin{equation}\label{eqn:ket_zero}
\left|\tilde{\lambda},0\right\rangle = \left(0,...,0,\left|0\right\rangle\right)^{\rm T}.
\end{equation}
where the notation is meant to indicate that $\lambda$ is ill defined due to an admixture of more than one band to form the $n=0$ Landau level.  With a single unique wavefunction, there can only be one Landau level with $n=0$.  Moving on, the $n=1$ spinor has two degrees of freedom found in its final two elements.  This allows for only two orthogonal constructions of the $n=1$ wavefunctions.  In this fashion, at a given value of $n$, there are $n+1$ allowed Landau levels, making this matter of limited states only an issue for $n < 2s$.  In this region of $n<2s$, the limited-number states are referred to as being ``shared" between the $2s+1$ bands.  The eccentric character of these states comes from the fact that the first few spinor elements in Eq.~(\ref{eqn:ket}) are zero.  This has been discussed with respect to the single $n=0$ level in graphene, which gives rise to an anomalous absorption peak in the magneto-optics.\cite{Gusynin07PRL}  The $n=0$ and $n=1$ Landau levels in bilayer graphene are idiosyncratic in the same way, where the wavefunction is not spread over all components of the spinor.\cite{Mireles12}

A pattern arises allowing one to determine which levels within a band are excluded.  If $s$ is a whole number, then a zero-energy flat band exists.  For  $n<2s$, only even numbered states exist within the flat band, down to $n=0$.  This is seen for $s=2$ in Fig.~\ref{fig:bare}(a), where each red dot represents a Landau level, the absence of which is seen for $n=\left\{1,3\right\}$ in the flat band.  The remaining bands come into existence at each successive odd-numbered $n<2s$, starting with the outer-most bands.  For example, in $s=2$, the bands $\lambda=\pm 2$ cover $n=\left\{1,2,...\right\}$ and the $\lambda=\pm 1$ bands have $n=\left\{3,4,...\right\}$.  In the absence of a flat band, i.e., when $2s+1$ is even, each successive pair of bands start on the even values of $n$.  The first Landau level in each band will be at an energy of zero and shared with its reflected band, as seen in Fig.~\ref{fig:bare}(b) for $s=\frac{3}{2}$.  This zero-energy state is a single Landau level shared between bands, not a case of two-fold Landau degeneracy.

The optical conductivity tensor is found via the Kubo formula.\cite{Mahan81}  In the Landau-level basis, this takes the form\cite{Tse11}
\begin{equation}\label{eqn:Kubo}
\sigma_{\alpha\beta}\left(\Omega\right) = \frac{ig}{2\pi\hbar l_{B}^{2}} \sum_{\rm LL's} \frac{n_{f} - n_{f}'}{\epsilon'-\epsilon} \frac{ \left\langle\psi\right|\hat{\jmath}_{\alpha}\left|\psi'\right\rangle \left\langle\psi'\right|\hat{\jmath}_{\beta}\left|\psi\right\rangle }{\Omega-\left(\epsilon'-\epsilon\right)+i\Gamma}.
\end{equation}
In Eq.~(\ref{eqn:Kubo}), the summation takes place over all initial and final Landau states, $\left|\psi\right\rangle\equiv\left|\lambda,n\right\rangle$ and $\left|\psi'\right\rangle\equiv\left|\lambda',n'\right\rangle$, respectively; $\Omega=h\nu$ is the photon energy; $n_{f}$ is the Fermi factor at a chemical potential $\mu$; $\Gamma$ is the scattering rate of charge carriers; and $\hat{\jmath}_{\alpha}=ev_{c}S_{\alpha}$ is the current operator, where $\alpha=\left\{x,y\right\}$.

In the limits of zero temperature and zero scattering rate, and using Eq.~(\ref{eqn:ket}), the following is obtained for the absorptive parts of the diagonal and off-diagonal components of the tensor:
\begin{equation}\label{eqn:sigmaxx,xy}
\begin{split}
\left.
\begin{array}{r}
 \mathrm{Re}\: \sigma_{xx} \left(\Omega\right)\\
 \mathrm{Im}\: \sigma_{xy} \left(\Omega\right)
\end{array}
\right\} =
& \frac{ge^{2}v_{c}^2}{8\hbar l_{B}^2}\sum_{\rm LL's} \frac{\delta\left(\Omega-\left(\epsilon'-\epsilon\right)\right)}{\epsilon'-\epsilon} \\
& \times\left(\left|f\left(\psi,\psi'\right)\right|^{2}\delta_{n',n-1} \pm \left|f\left(\psi',\psi\right)\right|^{2}\delta_{n',n+1}\right) \\
& \times\left[\theta\left(\mu-\epsilon\right)-\theta\left(\mu-\epsilon'\right)\right].
\end{split}
\end{equation}
The first line in this equation places sharp peaks in the spectrum located at the energy differences between two states, the height of each peak is reduced by said difference.  The final line, involving the Heaviside step function $\theta\left(x\right)$, ensures that a transition will only contribute to the spectrum if $\epsilon\leq\mu\leq\epsilon'$.  Note that at finite temperature, the step functions must be replaced by the appropriate Fermi factor $n_{f}$.  The middle line dictates the amplitude of a transition in addition to specifying the selection rule $n\rightarrow n\pm1$.  In this middle line, the plus sign is for $\rm{Re}\:\sigma_{xx}$ and the minus for $\rm{Im}\:\sigma_{xy}$.  The overlap function is given as
\begin{equation}\label{eqn:overlap}
f\left(\psi,\psi'\right)=\sum_{m=1}^{2s}\rho_{m}\check{\alpha}_{\lambda n}^{m}\alpha_{\lambda' n'}^{m+1},
\end{equation}
where $\check{x}$ denotes the complex conjugate of $x$.  Explicit expressions for $\left|f(\psi',\psi)\right|^{2}$ are given in Table \ref{tab:overlap} for $s=\frac{1}{2}$ and $s=1$.

\renewcommand{\arraystretch}{1.5}
\begin{table}
\caption{\label{tab:overlap}Explicit expressions for the relevant overlap $\left|f(\psi',\psi)\right|^{2}$ found in Eq.~(\ref{eqn:overlap}) for the $s=1/2$ and $s=1$ DW systems.}
\centering
\begin{tabular}{c l l}
\hline\hline
$\;s\;$ & $\;(\psi,\psi')$ & $\left|f(\psi',\psi)\right|^{2}$ \\
\hline
$\;\frac{1}{2}\;$ & $\left(\left|\tilde{\lambda},0\right\rangle,\left|\pm\frac{1}{2},1\right\rangle\right) $ & $\frac{1}{2}$ \\
 & $\left(\left|\lambda,n>0\right\rangle,\left|\pm\lambda,n+1\right\rangle\right)$ & $\frac{1}{4}$ \\
$\;1\;$ & $\left(\left|0,0\right\rangle,\left|\pm 1,1\right\rangle\right)$ & $1$ \\
 & $\left(\left|\pm 1,n\right\rangle,\left|0,n+1\right\rangle\right)$ & $\frac{n+1}{2n+1}$ \\
 & $\left(\left|0,n>1\right\rangle,\left|\pm 1,n+1\right\rangle\right)$ & $\frac{n-1}{2n-1}$ \\
 & $\left(\left|\lambda\neq 0,n\right\rangle,\left|\pm\lambda,n+1\right\rangle\right)$ & $\frac{n\left(2n\pm\sqrt{4n^{2}-1}\right)}{4n^{2}-1}$ \\
\hline\hline
\end{tabular}
\end{table}

Note that $\left|\Delta\lambda\right|>1$ transitions are strictly forbidden in the case of zero magnetic field.  This hard selection rule comes from the current operators, $\hat{\jmath}_{\alpha}\propto S_{\alpha}$, with the fact that $\lambda$ is a good quantum number in the $B=0$ case.  To see this, consider the interband transition matrix element
\begin{equation}
{\vphantom{\left\langle\psi\right\rangle}}_{0}\!\left\langle\psi\right|S_{x}\left|\psi'\right\rangle_{0} = {\vphantom{\left\langle\psi\right\rangle}}_{0}\!\left\langle\psi\right|\left(S_{+} + iS_{-}\right)\left|\psi'\right\rangle_{0}.
\end{equation}
$\left|\psi\right\rangle_{0}$ (where the subscript indicates the zero field wavefunction) must have $\lambda=\lambda'\pm 1$ for a non-zero matrix element.  In the presence of a field, however, there is a finite probability that $\left|\Delta\lambda\right|>1$ transitions will occur, as $\lambda$ is no longer considered a good quantum number.  It can be shown explicitly, however, that for each of the cases studied, the original selection rules and overall spectra\cite{Dora11} are recovered in the limit $B\rightarrow 0$.  Those transitions that are forbidden when $B=0$ have weak amplitudes in the finite-$B$ case, seen in Sec. \ref{sec:results}.

The form of the wavefunctions (derived in the appendix) is such that, for nonzero $\lambda$, $\left|\psi\right\rangle=\left|\lambda,n\right\rangle$ and $\left|\xi\right\rangle=\left|-\lambda,n\right\rangle$ have the same spinor structure except for a difference in sign of the even elements.  This leads to the following useful identity,
\begin{equation}\label{eqn:overlap_identity}
f\left(\psi,\psi'\right) =-f\left(\xi,\xi'\right),
\end{equation}
with $\left|\psi'\right\rangle=\left|\lambda',n\right\rangle$ and $\left|\xi'\right\rangle=\left|-\lambda',n\right\rangle$.  Furthermore, a state $\left|\psi_{n}\right\rangle=\left|\lambda,n\right\rangle$ is similar in structure to $\left|\psi_{m}\right\rangle$ for $n,m\geq 2s$ (Appendix \ref{sec:Alphas}).  That is, away from the region of shared states, the wavefunctions of Landau levels within each band are akin.  Then, for example, writing $\left|\psi'_{n}\right\rangle=\left|\lambda',n\right\rangle$, $f\left(\psi_{n+1},\psi'_{n}\right)$ will be a similar overlap to $f\left(\psi_{m+1},\psi'_{m}\right)$.  This gives the same character to peaks in a spectrum that come from transitions that are related in this way.  The similarity is not always true for transitions involving $n<2s$, as the wavefunctions here are dissimilar to those at higher $n$, arising in peaks with irregular-seeming amplitudes when compared to others.

As seen in graphene,\cite{Orlita10,Sadowski07} the optical conductivity for the absorbing part of circular-polarized light, $\rm{Re}\:\sigma_{\pm}=\rm{Re}\:\sigma_{xx}\mp\rm{Im}\:\sigma_{xy}$, involves only $n\rightarrow n+1$ or $n\rightarrow n-1$ transitions for right or left polarizations:
\begin{equation}\label{eqn:sigma_plus}
\begin{split}
\mathrm{Re}\:\sigma_{+}\left(\Omega\right)=\frac{ge^{2}v_{c}^{2}}{4\hbar l_{B}^{2}}\sum_{\rm LL's}&\frac{\delta\left(\Omega-\left(\epsilon'-\epsilon\right)\right)}{\epsilon'-\epsilon} \\
&\times\left|f\left(\psi',\psi\right)\right|^{2}\delta_{n',n+1} \\
&\times\left[\theta\left(\mu-\epsilon\right)-\theta\left(\mu-\epsilon'\right)\right],
\end{split}
\end{equation}
\begin{equation}\label{eqn:sigma_minus}
\begin{split}
\mathrm{Re}\:\sigma_{-}\left(\Omega\right)=\frac{ge^{2}v_{c}^{2}}{4\hbar l_{B}^{2}}\sum_{\rm LL's}&\frac{\delta\left(\Omega-\left(\epsilon'-\epsilon\right)\right)}{\epsilon'-\epsilon} \\
&\times\left|f\left(\psi,\psi'\right)\right|^{2}\delta_{n',n-1} \\
&\times\left[\theta\left(\mu-\epsilon\right)-\theta\left(\mu-\epsilon'\right)\right].
\end{split}
\end{equation}

In the integer-$s$ case, the partially filled flat band could give rise to important interactions both in the presence and absence of a magnetic field.  This could also occur at finite $\mu$ if a single Landau level is partially filled, however the effect may be more significant in the former case with the large density of states found in the flat band.  These interactions could change the ground state, potentially lifting the huge degeneracy.  However, to first approximation, we ignore such effects here.  In principle, it is possible to couple optically to the finite-$q$ plasmons through use of gratings,\cite{Mak12} nanoscopy,\cite{Fei11} or other forms of spatial confinement\cite{Yan12} as seen in graphene, or to access them through their effects via charge impurities.\cite{Kechedzhi13}  Such considerations are beyond the scope of this work and have not been included.  We also ignore other possibly relevant effects such as those that arise from plasmarons,\cite{Carbotte12} phonons,\cite{Pound12,Pound11a,Pound11b} spin-orbit interactions,\cite{Konschuh11,Tabert13PRB,Tabert13PRL} etc. which have not played a significant role in the magneto-optics of graphene ($s=\frac{1}{2}$).  Furthermore, we note a recent work by Orlita {\em et al}. on the three-dimensional zinc-blende crystal Hg$_{1-x}$Cd$_{x}$Te (MCT).\cite{Orlita14}  At a critical ratio of $x=x_{c}\approx 0.17$, charge carriers in MCT are termed ``Kane fermions," exhibiting a low-energy dispersion with two Dirac cones and an essentially flat band (as in our Fig.~\ref{fig:cones}, top right).  The bottom cone and the flat band are both filled, while the top cone is empty.  The optical conductivity of the Kane fermions was measured both in the presence and the absence of a magnetic field.  These spectra show similar features to the $s=1$ results shown here in Sec. \ref{sec:results_s1}, but modified for a third spatial dimension.\cite{Ashby13}  Their findings validate our approach to exclude these extra effects (plasmons, phonons, etc.) in order to find the primary features of the optical spectra.

\section{Results}
\label{sec:results}

In this section, we discuss the spectra obtained via Eqs.~(\ref{eqn:sigmaxx,xy}), (\ref{eqn:sigma_plus}), and (\ref{eqn:sigma_minus}) for the first four relevant values of $s$.  In each system studied, at least three different positive values of the chemical potential were considered.  If we denote the three lowest positive Landau energies as $0<\epsilon_{a} < \epsilon_{b} < \epsilon_{c}$, respectively, then the three values of $\mu$ taken satisfy $0 < \mu_{a} < \epsilon_{a}$, $\epsilon_{a}<\mu_{b}<\epsilon_{b}$, and $\epsilon_{b}<\mu_{c}<\epsilon_{c}$, respectively.  Comparison across these values shows how a spectrum will change as individual Landau levels become occupied.  When a level is filled, it can no longer be the final state in a transition, however it can now act as an initial state.  This, of course, is the Pauli exclusion principle in action.

As certain states go from being empty to filled, the explanation of the patterns seen in the spectrum are greatly aided by what we will refer to as ``snowshoe" diagrams (see Fig.~\ref{fig:snowshoe1_2}).  These are not to be confused with fan diagrams found in related work.\cite{Sadowski06}  In the construction of a snowshoe diagram, a horizontal line is first placed at $\epsilon=\mu$ in the ``unlaced''  diagram (a plot of $\epsilon_{\lambda n}$ vs. $n$ seen, for example, in Fig.~\ref{fig:bare}).  Next, arrows are drawn from all states below this line to any final state above it allowed by the selection rule $n\rightarrow n\pm 1$. The vertical component of each arrow is a measure of the energy difference between states. Furthermore, these arrows are color-coded to indicate intraband transitions ($\Delta\lambda=0$, blue), nearest interband transitions ($\Delta\lambda=\pm 1$, red), next-nearest interband transitions ($\Delta\lambda=\pm 2$, green), and so on.

With $\lambda$ no longer a good quantum number for finite $B$, some states exist that are not completely characteristic of a particular band.  Recall that for a particular $n<2s$, there are fewer states than there are bands.  Each state in this collection can be thought of as existing as shared across the $2s+1$ bands.  These special states give rise to transition peaks of irregular height.  The anomalous transitions should really be referred to as of a mixed-type instead of a specified inter- or intraband transition.  Despite this, the color-coding outlined above is still retained for the mixed-types.  The patterns that these peaks show during a shift in chemical potential are grouped well with the transitions of the same assigned color.  It will be seen that these patterns are dictated by the symmetry of a peak's representative arrow in the snowshoe diagram.

\subsection{$s=\frac{1}{2}$}

\begin{figure}
\begin{center}
\includegraphics[width=1.0\linewidth]{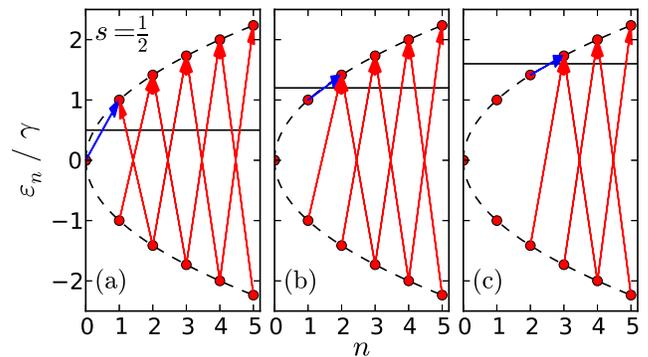}
\end{center}
\caption{\label{fig:snowshoe1_2}(Color online) Snowshoe diagrams for an $s=\frac{1}{2}$, or graphene-like DW at three values for the chemical potential: (a)$\mu_{a}=0.5\gamma$, (b)$\mu_{b}=1.2\gamma$, and (c)$\mu_{c}=1.6\gamma$, each indicated by a horizontal black line.  Blue arrows show allowed intraband transitions (including the mixed-type transition in panel (a)), while red show allowed interband transitions between Landau levels.}
\end{figure}

\begin{figure}
\includegraphics[width=\linewidth]{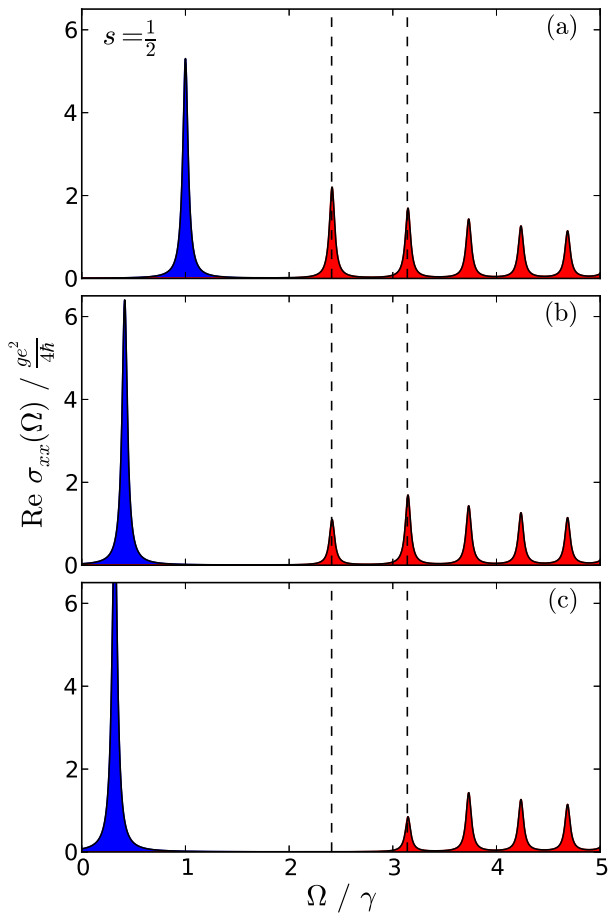}
\caption{\label{fig:sigmaxx1_2}(Color online) Absorptive diagonal component of the optical conductivity tensor for an $s=\frac{1}{2}$ Dirac-Weyl for three values of the chemical potential: (a)$\mu_{a}=0.5\gamma$, (b)$\mu_{b}=1.2\gamma$, and (c)$\mu_{c}=1.6\gamma$.  Blue peaks correspond to intraband transitions and red to interband.  Note that the blue peak in panel (a) is not strictly from an intraband transition, but is more aptly referred to as a mixed type.  Vertical dashed lines are placed at the photon energies $\Omega=2.41\gamma$ and $\Omega=3.14\gamma$.}
\end{figure}

\begin{figure}
\includegraphics[width=\linewidth]{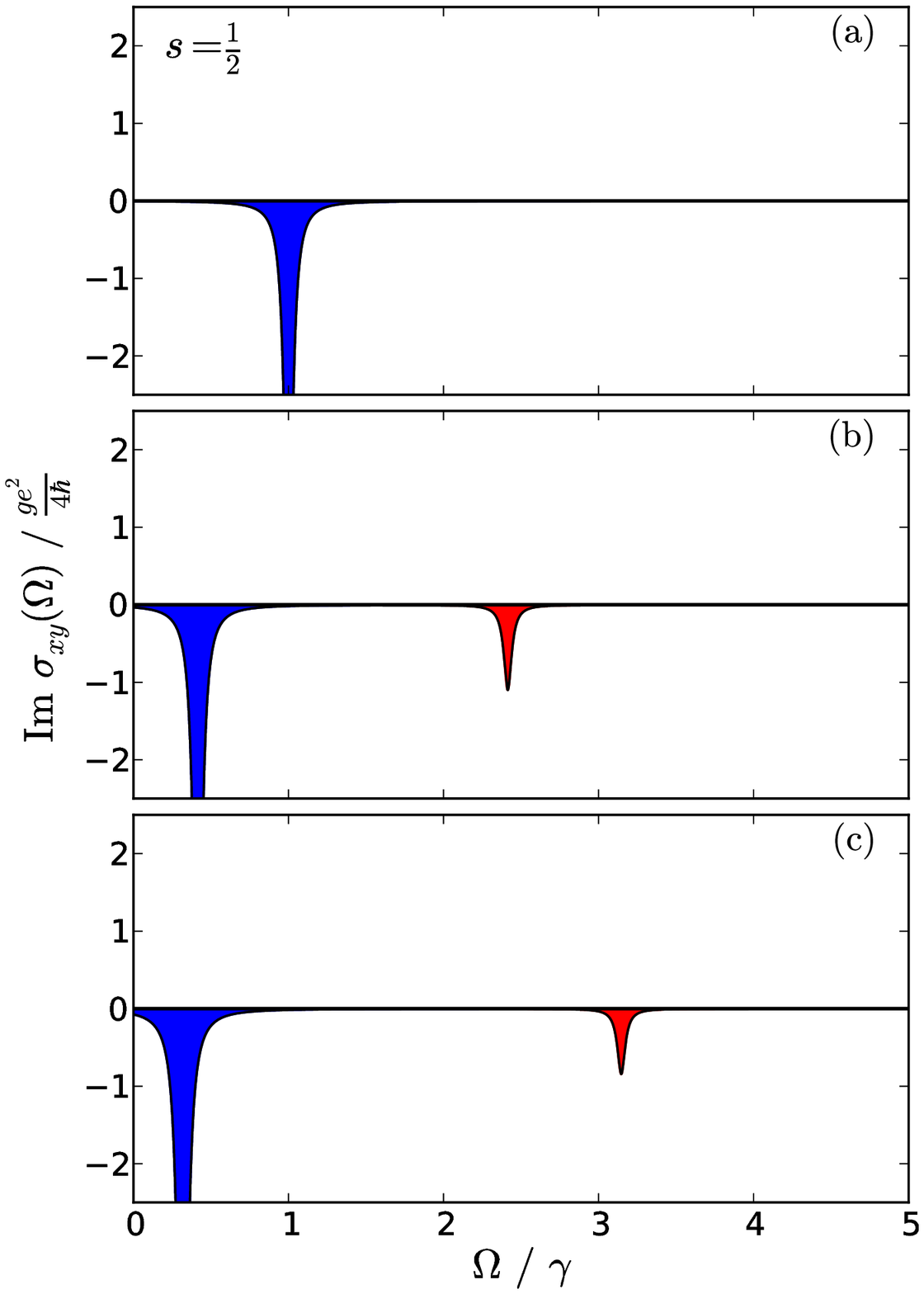}
\caption{\label{fig:sigmaxy1_2}(Color online) Absorptive off-diagonal component of the optical conductivity tensor for an $s=\frac{1}{2}$ Dirac-Weyl for three values of the chemical potential.  Labels are as in Fig.~\ref{fig:sigmaxx1_2}.}
\end{figure}

\begin{figure}
\includegraphics[width=\linewidth]{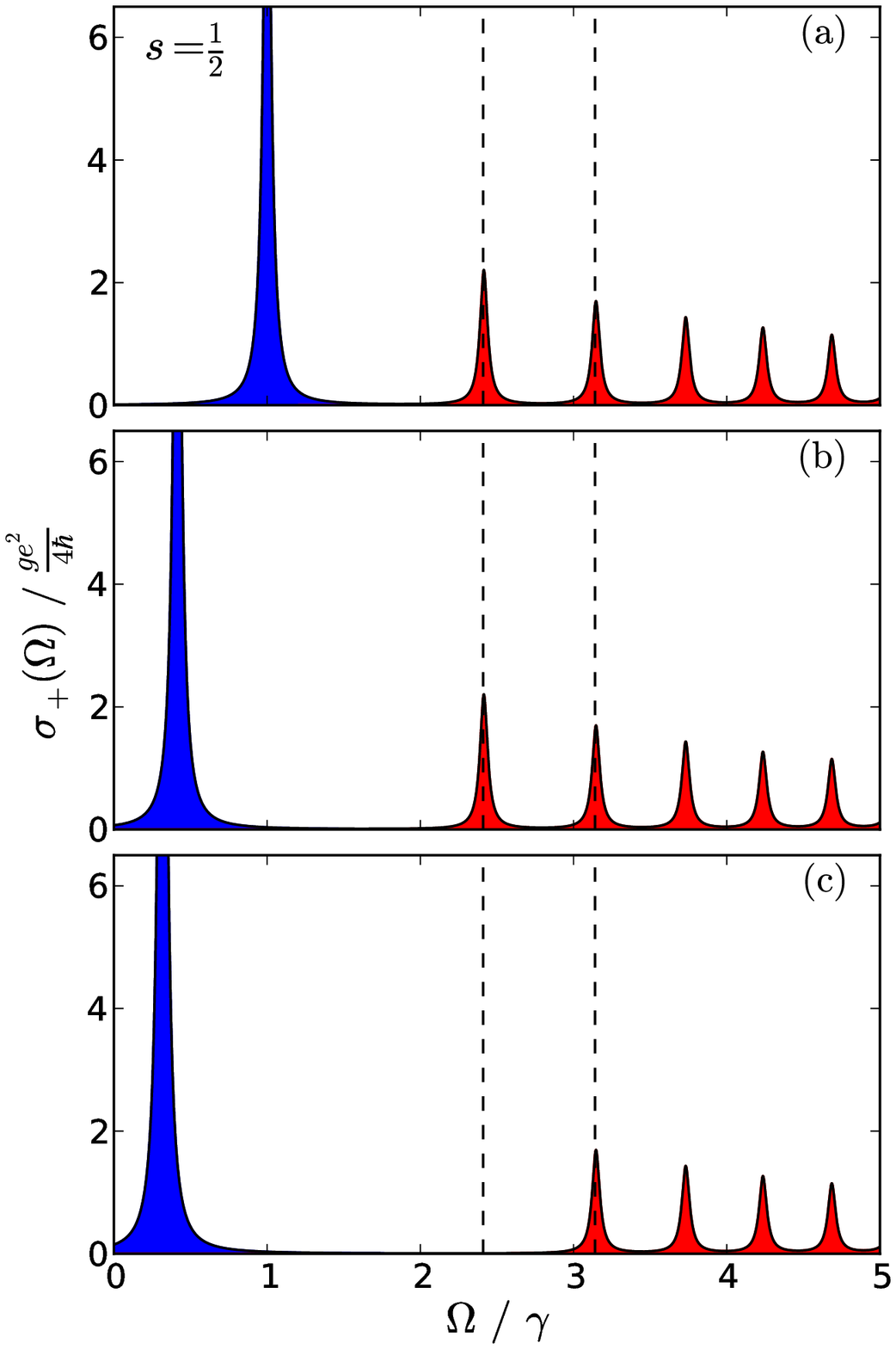}
\caption{\label{fig:sigmaplus1_2}(Color online) Absorptive optical conductivity for right-hand polarized light of an $s=\frac{1}{2}$ Dirac-Weyl for three values of the chemical potential.  Labels are as in Fig.~\ref{fig:sigmaxx1_2}.}
\end{figure}

\begin{figure}
\includegraphics[width=\linewidth]{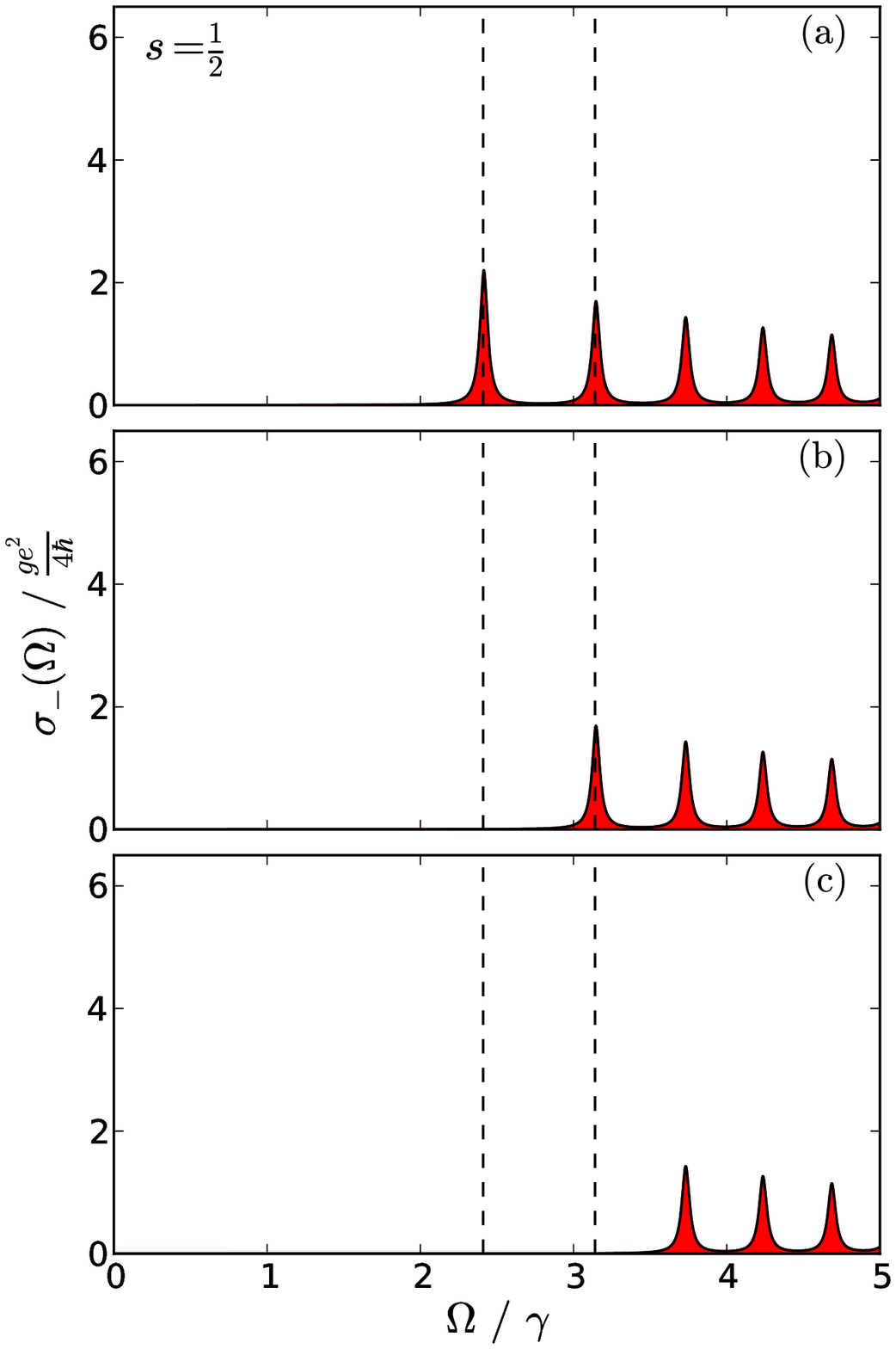}
\caption{\label{fig:sigmaminus1_2}(Color online) Absorptive optical conductivity for left-hand polarized light of an $s=\frac{1}{2}$ Dirac-Weyl for three values of the chemical potential.  Only interband transitions contribute.  Labels are as in Fig.~\ref{fig:sigmaxx1_2}.}
\end{figure}

While $s=\frac{1}{2}$ has been discussed previously in the literature,\cite{Pound12,Gusynin07JP} we rehearse the results within this subsection.  This allows us to introduce the snowshoe diagrams and provide plots for comparison to those of higher $s$ values.

Figure \ref{fig:snowshoe1_2} shows the snowshoe diagrams for three successive values of the chemical potential in a graphene-like ($s=\frac{1}{2}$) DW.  In each diagram there is only one allowed intraband transition, shown as a blue arrow, with the energy of these transitions decreasing as $\mu$ increases between diagrams.  In Fig.~\ref{fig:snowshoe1_2}(a), red arrows come in pairs that share the same energy.  If the left-directional arrow in a pair goes from $n\rightarrow n-1$, then its right-directional partner is $n-1\rightarrow n$.  For example, the red arrow going from $n=1$ to $n'=2$ and the arrow from $n=2$ to $n'=1$ each represent a transition with an energy difference of $\Delta\epsilon=2.41\gamma$.  Note that these pairs are mirror symmetric across the horizontal line $\epsilon=0$.  The identity of Eq.~(\ref{eqn:overlap_identity}) ensures that the overlap functions $f=(\psi,\psi')$ will be the same magnitude for each arrow in any mirror-symmetric pair like this. In the second diagram, the left-directional arrow in the first red pair disappears as its final state is now occupied.  This leaves a single unpaired arrow along with the remaining series of unchanged pairs.  Subsequently, in Fig.~\ref{fig:snowshoe1_2}(c), the two arrows pointing to the now-locked-out state have disappeared.  What remains is again a single unpaired arrow and a series of paired arrows.  This predictable pattern will continue as $\mu$ is shifted past consecutive Landau levels.

These patterns seen in the snowshoe diagrams translate to the patterns already established in the conductivity of graphene\cite{Gusynin07JP}.  The diagonal component, ${\rm Re}\:\sigma_{xx}$, is reproduced in Fig.~\ref{fig:sigmaxx1_2} using Eq.~(\ref{eqn:sigmaxx,xy}) for each chemical potential of interest. The idealized Dirac delta functions have been plotted instead as Lorentzians,
\begin{equation}
\delta\left(x\right)=\lim_{\Gamma\to 0}\frac{1}{\pi}\frac{\Gamma}{x^{2}+\Gamma^{2}}.
\end{equation}
In the plots, a value of $\Gamma=0.03\gamma$ was used for aesthetic reasons.  Contributions to the conductivity from intraband transitions are shown in blue and interbands shown in red.  However, the blue peak in Fig.~\ref{fig:sigmaxx1_2}(a) is better referred to as being from a mixed-type transition for it involves the $n=0$ state which is shared between bands.  In going from the first plot to the second, the single blue peak disappears and a new one with greater amplitude appears at a lower energy.  This is seen again in going from the second to the third diagram.  Each of these blue peaks maps from the blue arrow found in the corresponding snowshoe diagram.  The relative amplitude and position of these peaks come from the relative vertical length of their arrows.

Following down the first vertical dashed line in Fig.~\ref{fig:sigmaxx1_2}, the red peak is first halved and then disappears between diagrams.  Along the second vertical line, this peak is not halved until the second shift in $\mu$ and would further disappear in a fourth plot.  As $\mu$ is shifted, each red peak will first be halved, then disappear, in a cascading manner.  This can be understood by considering the snowshoe diagrams of Fig.~\ref{fig:snowshoe1_2}.  The full weight of the first red peak in Fig.~\ref{fig:sigmaxx1_2}(a) comes from the first pair of red arrows in Fig.~\ref{fig:snowshoe1_2}(a), where each arrow contributes an equal weight due to Eq.~(\ref{eqn:overlap_identity}).  These arrows represent the transitions $\left|-\frac{1}{2},1\right\rangle\rightarrow\left|+\frac{1}{2},2\right\rangle$ and $\left|-\frac{1}{2},2\right\rangle\rightarrow\left|+\frac{1}{2},1\right\rangle$.  As the chemical potential increases beyond the $\left|+\frac{1}{2},1\right\rangle$ level, Pauli exclusion forbids the latter transition.  This is seen in the next diagram where the arrow corresponding to this transition has disappeared, leading to an exact halving of the spectral peak.  As the chemical potential increases further, beyond the $\left|+\frac{1}{2},2\right\rangle$ level, the remaining transition is now forbidden.  The disappearance of the corresponding arrow at this energy in the third diagram illustrates the absence any peak.  Further, at this same moment the next peak in the series is halved, corresponding to the exclusion of the $\left|-\frac{1}{2},3\right\rangle\rightarrow\left|+\frac{1}{2},2\right\rangle$ transition.  This process of halving followed by elimination is the same for all pairs of red arrows and is triggered via the Pauli principle when a specific Landau level becomes occupied.  A significant feature of each of these interband peaks is that, when at half value, they represent only one transition of $n\rightarrow n+1$ ($\mu>0$) or $n\rightarrow n-1$ ($\mu<0$).  Then these halved peaks may be selected out by circularly polarized light as discussed below.

Figure \ref{fig:sigmaxy1_2} shows the off-diagonal component of the conductivity tensor, ${\rm Im}\:\sigma_{xy}$.  Due to the minus sign in the middle line of Eq.~(\ref{eqn:sigmaxx,xy}), mirror symmetric pairs of arrows in a snowshoe diagram cancel each other out in ${\rm Im}\:\sigma_{xy}$.  This is seen in Fig.~\ref{fig:sigmaxy1_2}(a), where there are no red peaks.  Each time $\mu$ is shifted past a Landau level, a single unpaired arrow remains.  These unpaired arrows produce peaks in the spectra, seen in Figs.~\ref{fig:sigmaxy1_2}(b) and \ref{fig:sigmaxy1_2}(c).  Similar to ${\rm Re}\:\sigma_{xx}$, the single intraband transition in each case (including the mixed-type transition at the lowest chemical potential) creates a blue peak at lower energies for an increasing chemical potential.  Thus, for graphene-like DW's, there will only be a maximum of two peaks in ${\rm Im}\:\sigma_{xy}$.  These peaks are negative for a positive chemical potential\cite{Gusynin07JP} which is related to the right-pointing direction of unpaired arrows in the snowshoe diagram.

The conductivity for right-hand polarized light, $\sigma_{+}$ shown in Fig.~\ref{fig:sigmaplus1_2}, has peaks corresponding to all right-directional arrows in the snowshoe diagrams, with a double-weight given to each arrow.  On the other hand, $\sigma_{-}$ shown in Fig.~\ref{fig:sigmaminus1_2} includes all left-directional peaks, doubly weighted.  With only a single arrow contributing to each peak, there is no halving of peak heights as a level becomes occupied.  Instead, each peak will in turn completely disappear.  The half-height peaks of Fig.~\ref{fig:sigmaxx1_2} will thus only show up in one circular polarization ($\sigma_{+}$ in this case) and may be accessed uniquely in this manner.

\subsection{$s=1$}
\label{sec:results_s1}

\begin{figure}
\includegraphics[width=\linewidth]{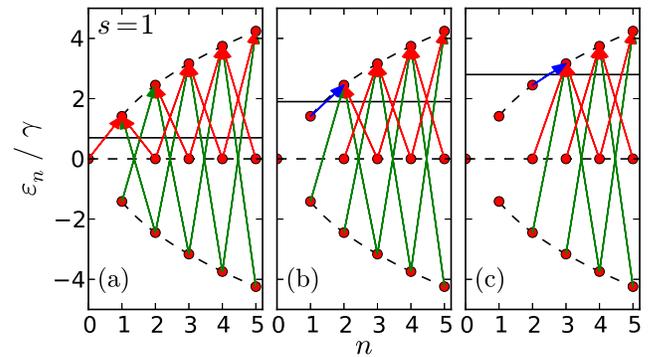}
\caption{\label{fig:snowshoe1}(Color online) Snowshoe diagrams for an $s=1$ DW at three values for the chemical potential: (a)$\mu_{a}=0.7\gamma$, (b)$\mu_{b}=1.9\gamma$, and (c)$\mu_{c}=2.8\gamma$, each indicated by a horizontal black line.  Blue arrows show allowed intraband transitions, red show nearest interband transitions, and green show next-nearest interband transitions between Landau levels.  Note that transitions involving $n<2$ are of a mixed type instead of having a strict intra- or interband classification.}
\end{figure}

\begin{figure}
\includegraphics[width=\linewidth]{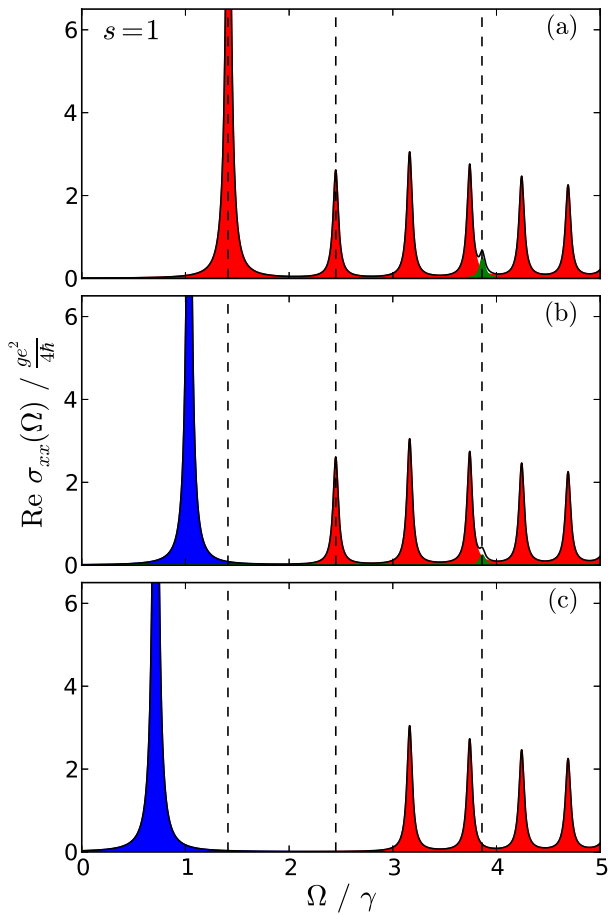}
\caption{\label{fig:sigmaxx1}(Color online) Absorptive diagonal component of the optical conductivity tensor for an $s=1$ Dirac-Weyl for three values of the chemical potential: (a)$\mu_{a}=0.7\gamma$, (b)$\mu_{b}=1.9\gamma$, and (c)$\mu_{c}=2.8\gamma$.  Blue peaks correspond to intraband transitions, red, to nearest interband, and green, to next-nearest interband.  Note that some peaks, such as the first red peak in panel (a), are of a mixed type instead of a strict intra- or interband type.  Vertical dashed lines are placed at the photon energies $\Omega=1.41\gamma$, $\Omega=2.45\gamma$, and $\Omega=3.86\gamma$.}
\end{figure}

\begin{figure}
\includegraphics[width=\linewidth]{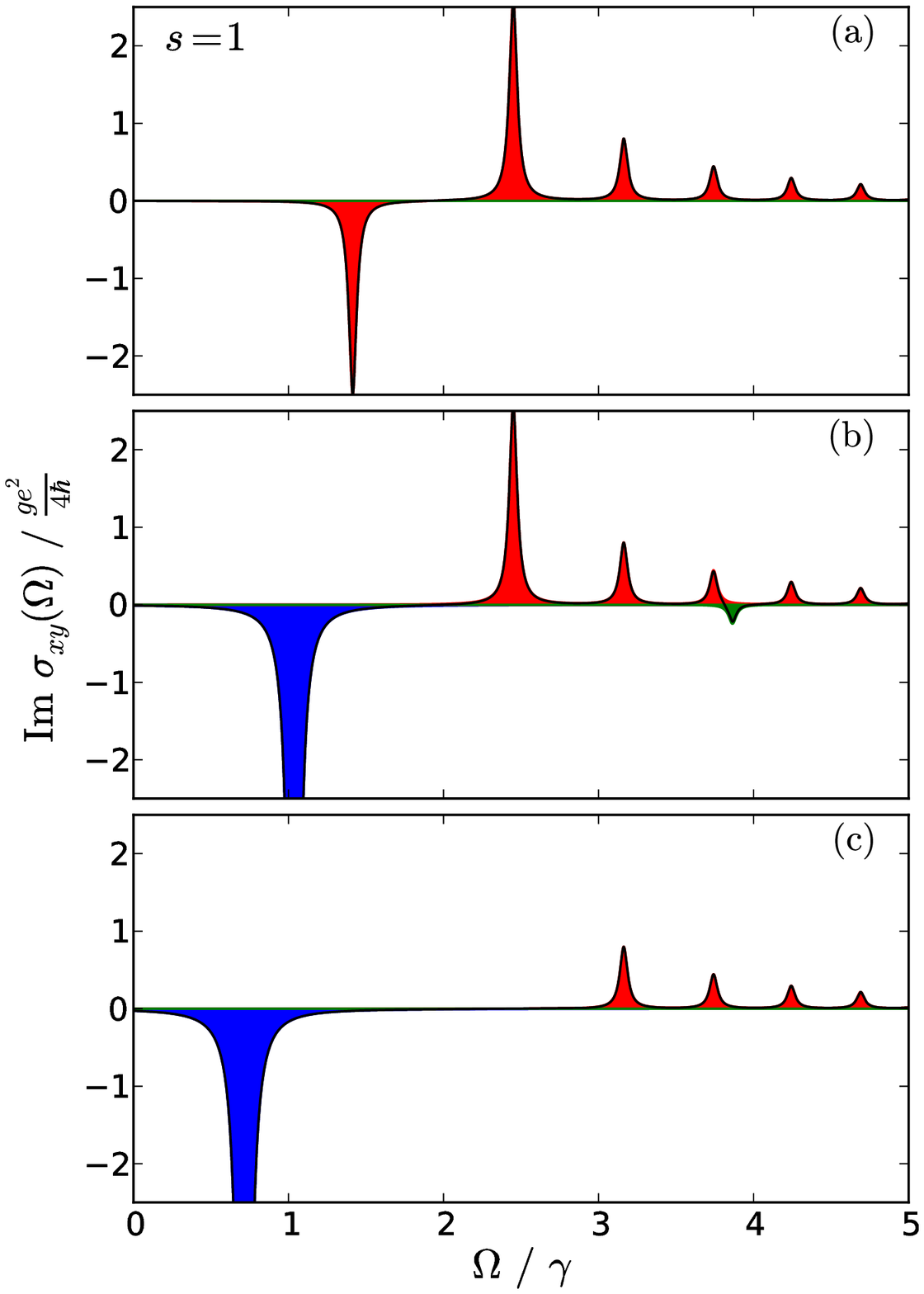}
\caption{\label{fig:sigmaxy1}(Color online) Absorptive off-diagonal component of the optical conductivity tensor for an $s=1$ Dirac-Weyl for three values of the chemical potential.  Labels are as in Fig.~\ref{fig:sigmaxx1}.}
\end{figure}

\begin{figure}
\includegraphics[width=\linewidth]{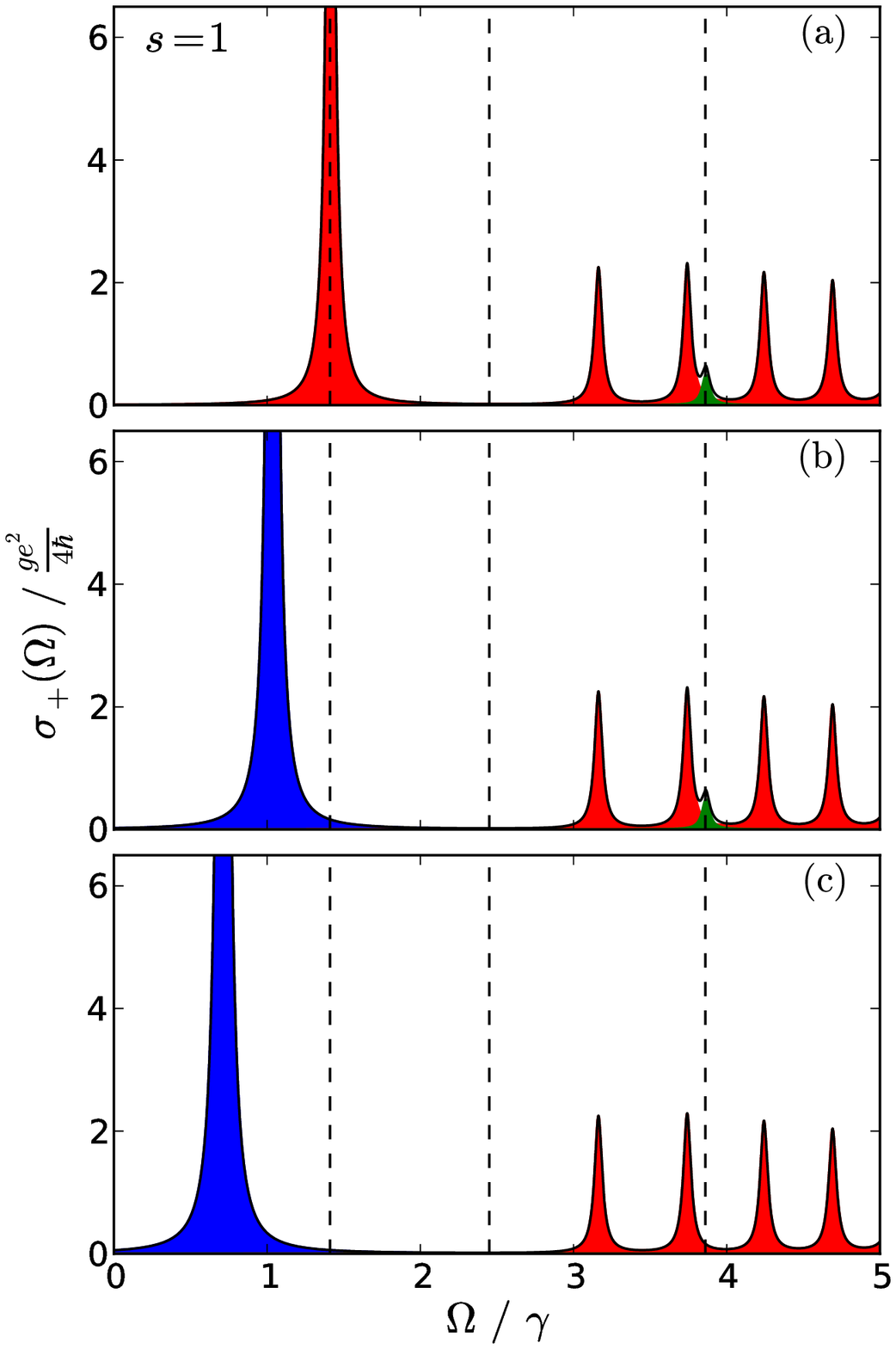}
\caption{\label{fig:sigmaplus1}(Color online) Absorptive optical conductivity for right-hand polarized light of an $s=1$ Dirac-Weyl for three values of the chemical potential.  Labels are as in Fig.~\ref{fig:sigmaxx1}.}
\end{figure}

\begin{figure}
\includegraphics[width=\linewidth]{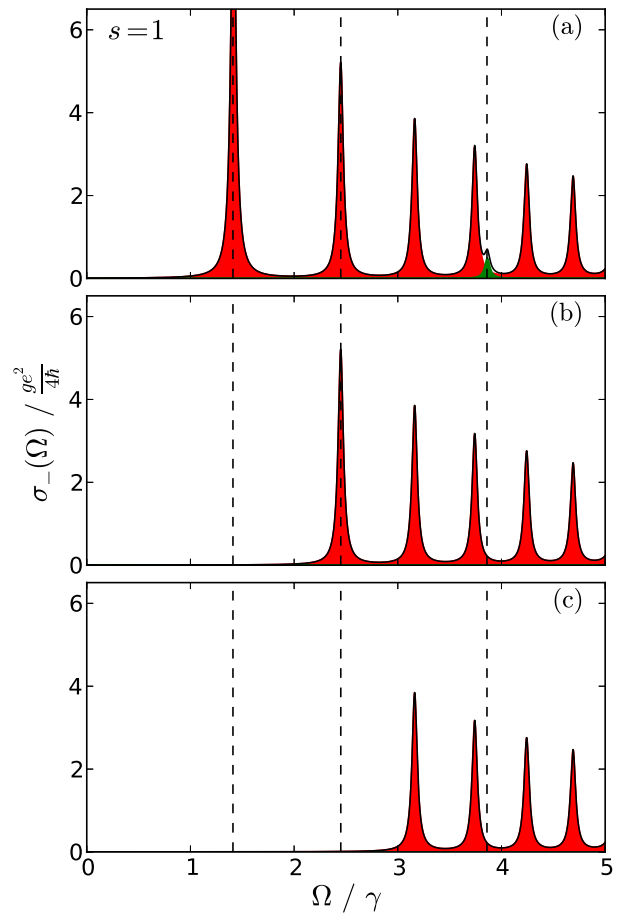}
\caption{\label{fig:sigmaminus1}(Color online) Absorptive optical conductivity for left-hand polarized light of an $s=1$ Dirac-Weyl for three values of the chemical potential.  Intraband transitions do not contribute.  Labels are as in Fig.~\ref{fig:sigmaxx1}.}
\end{figure}

The patterns seen in the spectra of $s=1$ DW's are noticeably different from those in the previous case, but are equally as predictable.  Figure \ref{fig:snowshoe1} has the $s=1$ snowshoe diagrams for the chemical potentials (a)$\mu_{a}=0.7\gamma$, (b)$\mu_{b}=1.9\gamma$, and (c)$\mu_{c}=2.8\gamma$.  One should make note of the the first red arrow in Fig.~\ref{fig:snowshoe1}(a), indicating a nearest interband transition from $\lambda=0$ to $\lambda=1$.  In truth, this is a mixed-type transition, as defined above.  All transitions involving states with $n<2s$ are of this class.

The green arrows come in mirror-symmetric pairs akin to those found in the graphene case.  As such, the peaks corresponding to these arrows show the same changes described above while $\mu$ is shifted across Landau levels.  The difference is that these peaks have a smaller amplitude and a larger interpeak spacing compared to the $s=\frac{1}{2}$ peaks.  While the red arrows also come in pairs of equal energy, they are not mirror-symmetric across $\epsilon=0$.  Instead, they form the arms of an isosceles triangle extending from $\epsilon=0$.  This mirror asymmetry means that their overlap functions are of different magnitude and thus do not cancel out in ${\rm Im}\:\sigma_{xy}$.

The $s=1$ diagonal component ${\rm Re}\:\sigma_{xx}$ is presented in Fig.~\ref{fig:sigmaxx1} for each of the three chemical potential values.  A green peak is seen along the third dashed line corresponding to the first pair of green arrows in Fig.~\ref{fig:snowshoe1}.  Between the spectra, the peak is first exactly halved and then disappears, showing the same pattern as the red peaks in the graphene case for mirror-symmetric pairs of arrows.  These $\Delta\lambda=2$ interband transitions in $s=1$ only produce minor features and present as small shoulders on larger peaks in the total conductivity.  The smallness of this feature is a remnant of the fact that such a transition is forbidden when $B=0$.  The two mixed-type transitions referred to in the previous paragraph produce the peak along the first dashed line.  It disappears in Fig.~\ref{fig:sigmaxx1}(b) as the final state in the transition has become occupied.  It is at the second value $\mu=\mu_{b}$ that we begin to see intraband transitions in the $\lambda=1$ band.  These transitions also follow the same patterns as the graphene intraband transitions.  Between subsequent plots, the blue peak disappears while a new one appears at lower energy with greater amplitude.  The spectra have a series of nearest interband peaks like those found in graphene, composed of two equal-energy snowshoe arrows.  However, in $s=1$, a shift in $\mu$ will block out both constituent transitions as they share the same final state.  As such, between each plot, the red peaks in the series take turns completely disappearing.  There is no intermediate step of finding the same peak with half the height.  This is a notable difference from the previous $s=\frac{1}{2}$ case.  A peculiarity is found along the second dashed line at an energy of $\Omega=2.45\gamma$, where the peak is contrastingly smaller in height than the one found to its right.  The absence of the $n=1$ state in the flat band means that there is only one transition with this energy.  With only one arrow contributing to the peak, it creates an irregularity in the steady decline of peak heights, appearing shorter than one might expect.  Thus, the $s=1$ spectrum at low doping (Fig.~\ref{fig:sigmaxx1}(a)) has the second peak lower in height than the third peak.  In comparison, this is another distinction from the $s=\frac{1}{2}$ spectrum Fig.~\ref{fig:sigmaxx1_2}(a), where the peak heights are monotonically decreasing.

In the off-diagonal spectrum (Fig.~\ref{fig:sigmaxy1}), the blue intraband peaks and green $\Delta\lambda=2$ interband peaks have the same features as their counterparts in graphene.  It is the nearest interband peaks in $s=1$ (those associated with the flat band) that provide a distinction from the $s=\frac{1}{2}$ spectrum.  That the pairs of equal-energy arrows do not possess mirror-symmetry means they do not cancel each other out here.  For each pair, except the one involving the mixed transition, the overlap function for the left-directional arrow is greater in magnitude than its right-directional partner.  The peculiarity from the unpaired arrow discussed in the previous paragraph now has the opposite effect in the off-diagonal conductivity.  Without a right-directional arrow to compete with, the unpaired arrow creates an irregularly large peak, the first positive peak in panels (a) and (b).  As such, peaks associated with each pair will be positive due to the $n\rightarrow n-1$ dominance.  The mirror asymmetry referred to in the snowshoe diagram creates a series of peaks in $\rm{Im}\:\sigma_{xy}$ not seen in the $s=\frac{1}{2}$ system.

Finally, Figs.~\ref{fig:sigmaplus1} and \ref{fig:sigmaminus1} give the optical conductivity for circular-polarized light.  Again, the blue and green peaks show graphene-like patterns under a changing chemical potential.  The series of red peaks will disappear one by one as $\mu$ is increased.  The absence of the $n=1$ Landau level in the flat band forces that there is no right-directional arrow with an energy of $\Omega=2.45\gamma$.  This means that there is no peak in the $\sigma_{+}$ plot at this energy, indicated by the second dashed line.  As such, the selection of transitions by either polarization has removed the non-monotonic decline of the peak heights seen in ${\rm Re}\:\sigma_{xx}$.  The relative difference in magnitude of the overlap functions of the equal-energy pairs of arrows causes red peaks in $\sigma_{-}$ to be larger than their counterparts in $\sigma_{+}$.  Note that the opposite is true for the peak along the first dashed line, which involves mixed-type transitions. One clear difference between this pair of spectra and their $s=\frac{1}{2}$ counterparts (Figs.~\ref{fig:sigmaplus1_2} and \ref{fig:sigmaminus1_2}) stands out.  The second peak in the series of interband peaks, indicated by the second dashed line in all graphs, has disappeared from the $s=1$ $\sigma_{+}$ spectrum.  Switching between circular polarizations gives a clear signature of either an $s=\frac{1}{2}$ or $s=1$ DW.

\begin{figure}
\includegraphics[width=\linewidth]{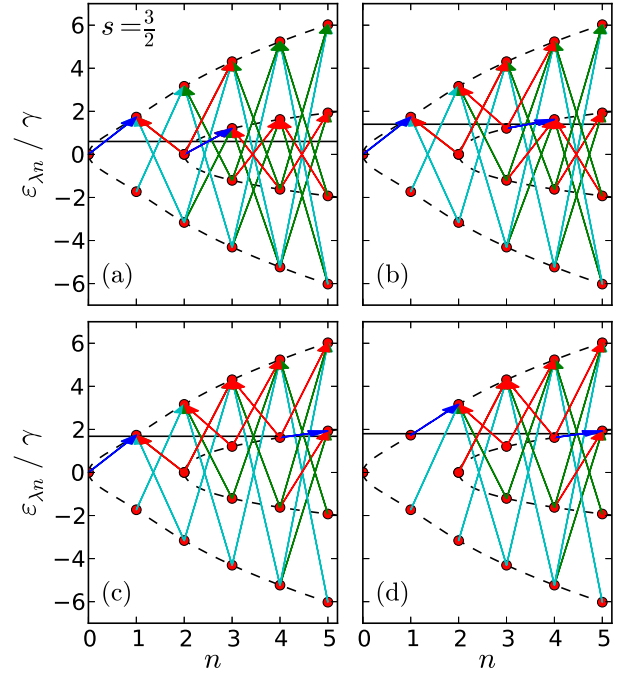}
\caption{\label{fig:snowshoe3_2}(Color online) Snowshoe diagrams for an $s=\frac{3}{2}$ DW at four values for the chemical potential: (a)$\mu_{a}=0.6\gamma$, (b)$\mu_{b}=1.4\gamma$, (c)$\mu_{c}=1.68\gamma$, and (d)$\mu_{d}=1.8\gamma$, each indicated by a horizontal black line.  Blue arrows show allowed intraband transitions, red show nearest interband transitions, green next-nearest interband, and cyan next-next-nearest interband transitions between Landau levels.  Note that transitions involving states with $n<3$ are of mixed-type.}
\end{figure}

\subsection{$s=\frac{3}{2}$}

\begin{figure}
\includegraphics[width=\linewidth]{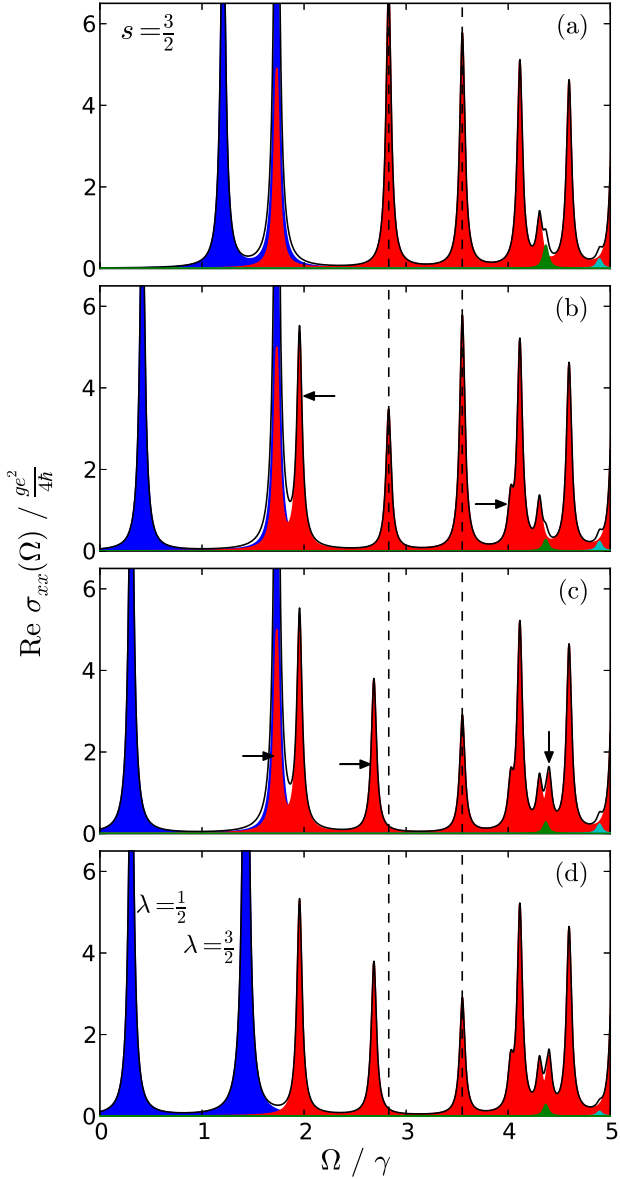}
\caption{\label{fig:sigmaxx3_2}(Colour online) Absorptive diagonal component of the optical conductivity tensor for an $s=\frac{3}{2}$ Dirac-Weyl for four values of the chemical potential: (a)$\mu_{a}=0.6\gamma$, (b)$\mu_{b}=1.4\gamma$, (c)$\mu_{c}=1.68\gamma$, and (d)$\mu_{d}=1.8\gamma$.  Blue peaks correspond to intraband transitions, red, to nearest interband, green, to next-nearest interband, and cyan to next-next-nearest interband.  Note that some peaks, such as the two blue peaks in panel (a), are of mixed-type.}
\end{figure}

As $s$ increases further, the optical conductivity becomes increasingly complicated.  However, its features can still be explained through snowshoe diagrams.  Figure \ref{fig:snowshoe3_2} gives the $s=\frac{3}{2}$ snowshoe diagrams for four positive values of the chemical potential.  A fourth value was included in order to observe the occupation of an empty state in the top band, seen between Figs.~\ref{fig:snowshoe3_2}(c) and \ref{fig:snowshoe3_2}(d).  In the last three diagrams, the proximity of the values of $\mu$ makes it difficult to discern a change in elevation of the horizontal black line.  A more illustrative indication of where the chemical potential lies comes from the fact that arrows only either begin or terminate at occupied or unoccupied states, respectively.  Again, transitions involving states with $n<2s$ are better defined as mixed-type despite the color assigned to them in Fig.~\ref{fig:snowshoe3_2}.

Figure \ref{fig:sigmaxx3_2} has the $s=\frac{3}{2}$ diagonal component ${\rm Re}\:\sigma_{xx}$ corresponding to each of the four snowshoe diagrams.  With two positive bands, there are now two possible intraband transitions for all finite values of $\mu$, one for $\left|\lambda\right|=\frac{1}{2}$ and another for $\left|\lambda\right|=\frac{3}{2}$, found at higher energy.  These two peaks are labelled in the bottom panel, Fig.~\ref{fig:sigmaxx3_2}(d).  The nearest interband transitions from $\lambda=-\frac{1}{2}\rightarrow\lambda'=\frac{1}{2}$ show the same snowshoe symmetry as those found in the $s=\frac{1}{2}$ snowshoe diagrams (Fig.~\ref{fig:snowshoe1_2}).  This leads to a series of red peaks that are first halved then disappear as certain Landau levels become occupied.  Vertical lines at photon energies $\Omega=2.83\gamma$ and $\Omega=3.55\gamma$ show two peaks that illustrate this.  Note that the half-height peak along the second dashed line does not disappear between the third and fourth plot.  This is because the shift in $\mu$ here does not change the occupation of the states involved in this transition.  Instead, this shift fills a state in the outer band, causing a disappearance of the compound blue and red peak at $\Omega=1.73\gamma$.  The extra band also creates an entire subset of nearest interband transitions between $\lambda=\frac{1}{2}$ and $\lambda'=\frac{3}{2}$.  This subset produces peaks that wholly appear and disappear without halving.  Specifically, when a certain initial state becomes occupied, two peaks appear at different energies.  As the chemical potential is raised, the lower energy peak will disappear first, followed by the higher energy one.  The complete asymmetry of the arrows found in this subset show that there are no equal-energy pairs.  The arrows in panel (b) and the two arrows on the right in panel (c) show the appearance of such pairs of peaks.  These peaks arise from $\lambda=\frac{1}{2}\rightarrow\frac{3}{2}$ transitions out of $n=3$ and $n=4$, respectively.  The left-most arrow in panel (c) indicates a peak of this type that disappears in the next shift of $\mu$.  Next- and next-next-nearest interband transitions come in mirror symmetric pairs that will follow the graphene halving/disappearance pattern.  However, the large separation in energy and band index in these transitions mean that they only produce small effects in the spectrum at energies higher than other interesting features.  In the energy range plotted, only two such peaks appear (green and cyan) that present as shoulders in the total spectrum.  These peaks correspond to the first mirror symmetric pair of arrows of each color in the first snowshoe diagram.  We see that the green peak is halved in height between the first and second panels, and the cyan peak is halved between the final two panels.  Eventually, by tuning the chemical potential further, these peaks will wholly disappear from the spectrum.

As seen previously, a plot of ${\rm Im}\:\sigma_{xy}$ for $s=\frac{3}{2}$ will show peaks corresponding to any arrows that are not cancelled out by a mirror-symmetric partner in the snowshoe diagram.  These peaks will be positive for left-directional arrows and negative for right-directional arrows.  In the case of competing left and right arrows, as in the mixed transitions with final state $\left|\frac{3}{2},1\right\rangle$, the arrow with the larger overlap function will dominate in the spectrum (the blue arrow in this particular pair).  For cirular polarization, $\sigma_{+}$ shows all peaks due to right-directional arrows and $\sigma_{-}$ all those due to arrows pointing to the left.  For brevity, these plots are not shown here.

\subsection{$s=2$}

\begin{figure}
\includegraphics[width=\linewidth]{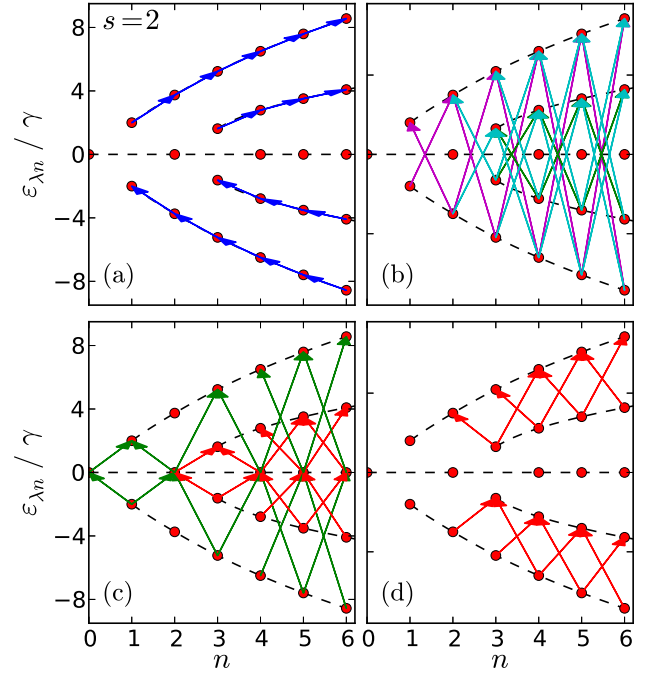}
\caption{\label{fig:snowshoe2}(Color online) Snowshoe diagrams of an $s=2$ DW showing the four types of transitions.  These types are (a)intraband, (b)mirror-symmetric pairs, (c)flat-band transitions, and (d)asymmetric.  Note that for a given chemical potential, many of these transitions are locked out.  While each arrow in these illustrative diagrams has a symmetric partner reflected across $\epsilon=0$, this is only retained in panel (b) when a finite chemical potential is specified.}
\end{figure}

\begin{figure}
\includegraphics[width=\linewidth]{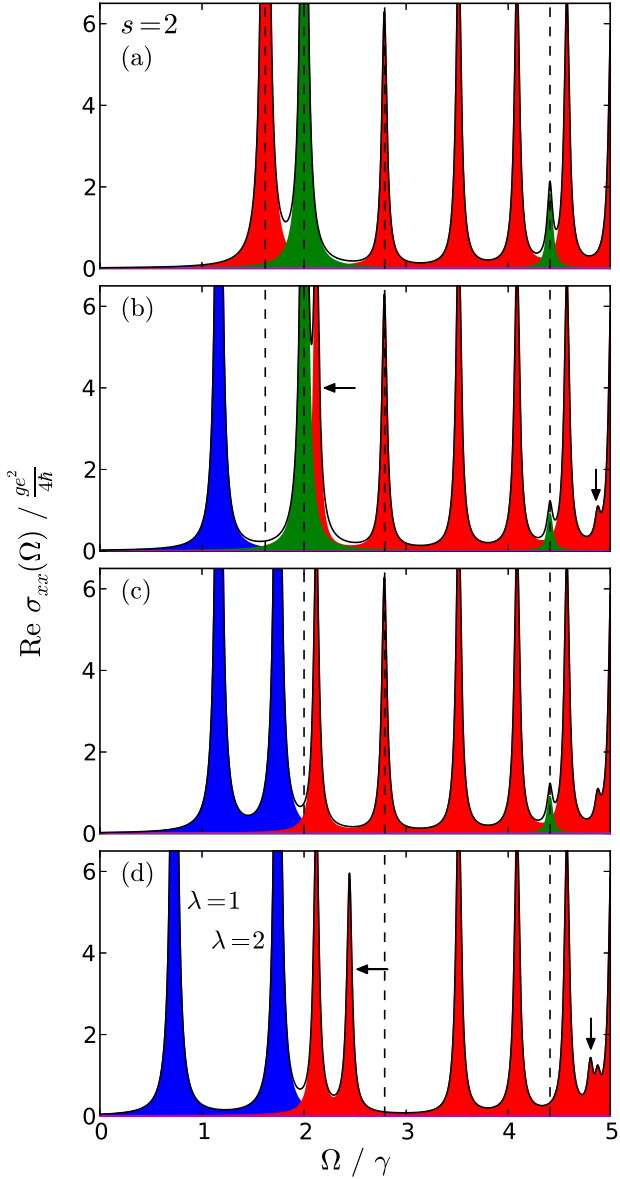}
\caption{\label{fig:sigmaxx2}(Color online) Diagonal component of the optical conductivity tensor for an $s=2$ Dirac-Weyl for four values of the chemical potential: (a)$\mu_{a}=0.8\gamma$, (b)$\mu_{b}=1.8\gamma$, (c)$\mu_{c}=2.4\gamma$, and (d)$\mu_{d}=3.2\gamma$.  Blue peaks correspond to intraband transitions, red to nearest interband, and green to next-nearest interband.}
\end{figure}

In the $s=2$ DW, peaks in the optical spectra can be grouped into four families which determine their characteristics with a change in $\mu$.  Figure \ref{fig:snowshoe2} groups all possible transitions in the $s=2$ snowshoe diagrams into the four families.  These diagrams ignore the Pauli exclusion principle and the role of $\mu$, so that for a particular chemical potential many of these transitions will be locked out.  Figure \ref{fig:sigmaxx2} gives the $s=2$ ${\rm Re}\:\sigma_{xx}$ spectrum for four successive values of $\mu$.

Fig.~\ref{fig:snowshoe2}(a) shows all of the intraband transitions, of which there are none in the flat band.  When imposing the exclusion principle, only a maximum of two intraband transitions are possible.  These will be in the bands $\left|\lambda\right|=1$ and $\left|\lambda\right|=2$, respectively, with $\sgn\lambda=\sgn\mu$.  The two intraband peaks in Fig.~\ref{fig:sigmaxx2}(d) are labelled by their band index.  As $\mu$ shifts higher, an intraband peak may wholly disappear with another taking its place at lower energy and with greater amplitude.

Next are those transitions analogous to the $s=\frac{1}{2}$ interband transitions, displayed in Fig.~\ref{fig:snowshoe2}(b).  When the exclusion principle is imposed, these transitions appear in mirror-symmetric pairs of arrows across $\epsilon=0$ in the snowshoe diagram.  For $s=2$, we see that there are three subgroups within this family: next-nearest (green arrows), next-next-nearest (cyan), and ${\rm N}^{3}$-nearest (magenta) transitions.  Note again that transitions involving a state with $n<2s$, or $n<4$ here, should be referred to as mixed-type transitions.  Each subgroup produces a series of peaks in the spectrum with decreasing amplitude at higher energies.  As $\mu$ is increased, the left-most peak in a series may disappear while the peak to its right is reduced in height by one half.  This pattern continues successively for an ever increasing chemical potential.  For the limited energy scale in Fig.~\ref{fig:sigmaxx2}, only one of these peaks is present.  This is found at an energy of $\Omega=4.41\gamma$ and is indicated by the right-most dashed line in each panel.  Being the first peak in the series, there is no disappearing peak to its left when this one is halved between panels (a) and (b).  This halved peak then fully disappears between panels (c) and (d).

For the next family of transitions, Fig.~\ref{fig:snowshoe2}(c) fully includes all transitions from or to the flat band.  In this family, pairs of equal-energy arrows form the arms of an isosceles triangle with the base lying on $\epsilon=0$.  With the exclusion principle, we see that there are two subgroups: nearest (red) and next-nearest (green) interband transitions.  These translate into the spectrum as two series of peaks with generally decreasing amplitudes at higher energies.  Similar to $s=1$, we see an anomalous peak along the third dashed line in Fig.~\ref{fig:sigmaxx2}(a) that breaks the monotonic decline of peak amplitudes.  This is due to there being only a single transition out of the flat band with the final state $\left|1,4\right\rangle$, instead of two.  As $\mu$ increases, peaks will in turn wholly disappear starting with the left-most.  The nearest interband series produce all of the tall red peaks in Fig.~\ref{fig:sigmaxx2}(a), the first two of which are indicated by dashed lines.  The first peak wholly disappears between panels (a) and (b), while the second remains unchanged.  The second peak then disappears as $\mu$ is shifted between panels (c) and (d), leaving all other peaks in this series unchanged.  The first next-nearest peak in the family is found in panels (a) and (b) as the large green peak at an energy $\Omega=2\gamma$.  This peak should really be referred to as a mixed-type instead of next-nearest.  Whatever the classification, however, this peak still belongs in the subgroup with all other next-nearest transitions from the flat band.  The next peak in this series is at $\Omega=5.23\gamma$ and does not appear in the plot.  Referring back to the first peak, it has wholly disappeared between panels (b) and (c).  Note that this pattern is exactly what was seen in $s=1$, the other DW with a flat band studied.  The difference between cases is that a second series of peaks arises from a second positive energy band found in $s=2$.

The final family of transitions, drawn in Fig.~\ref{fig:snowshoe2}(d), collect what is left over.  These consist of transitions between bands $\lambda$ and $\lambda'\neq\lambda$, where $\sgn\lambda=\sgn\lambda'$ and $\lambda,\lambda'\neq 0$.  Such transitions were found in $s=\frac{3}{2}$ with $\lambda=\frac{1}{2}$ and $\lambda'=\frac{3}{2}$.  Within this family, as $\mu$ is shifted, arrows appear or disappear in the snowshoe diagram without any symmetric pairs.  As such, the spectra will see individual peaks appear and disappear.  Arrows in Fig.~\ref{fig:sigmaxx2} point to the appearance of peaks like this.  Referring to the peak in panel (b) indicated by the arrow on the left, its disappearance would occur when $\mu$ is shifted through the value $\mu=3.74\gamma$.  The other peaks would disappear at higher values of $\mu$, while new peaks of this type may occur by tuning $\mu$.

The spectra for other polarizations can be explained or even constructed to a qualitative degree from the snowshoe diagrams.  For ${\rm Im}\:\sigma_{xy}$, one must only recall that arrows pointing to the right have a negative contribution to the spectrum.  Thus, any mirror-symmetric pairs of arrows in the snowshoe diagram have zero contribution and a competition exists between equal-energy pairs that are not mirror symmetric.  For the circularly polarized spectra, only arrows to the right contribute (positively) to ${\rm Re}\:\sigma_{+}$ and arrows to the left contribute to ${\rm Re}\:\sigma_{-}$.

\section{Conclusions}
\label{sec:conc}

Derivation of the optical conductivity for a system of DW fermions with arbitrary pseudospin in an external magnetic field resulted in the expressions given in Eqs.~(\ref{eqn:sigmaxx,xy}), (\ref{eqn:sigma_plus}), and (\ref{eqn:sigma_minus}).  The equations describe spectra with peaks representing transitions between Landau levels.  In $B=0$, one finds a strict interband selection rule of $\Delta\lambda=1$ for all systems discussed here.  However, with finite $B$, there is no restriction of $\Delta\lambda$.  Instead, we have the well-known rule $n\rightarrow n\pm 1$.  Nonetheless, an artefact of the former rule remains, as the $\Delta\lambda>1$ transitions have only weak amplitudes.  From the snowshoe diagrams, we see that transitions can be grouped into four families based on their symmetry that will provide a predictable pattern to optical conductivity spectra.  

(1) \textit{Intraband}:  Within a particular band in the snowshoe diagram, this transition will be between the highest occupied and lowest unoccupied Landau levels in the band.  As the chemical potential is shifted above the final state in this transition, it now becomes the initial state in a new intraband transition.  This shift causes one peak in the diagonal spectrum to disappear and a new one to appear at a lower energy with higher amplitude.  

(2) \textit{Mirror-symmetric pairs}: When snowshoe arrows come in pairs that are mirror-symmetric across $\epsilon=0$, they produce features like those found in the graphene interband transitions.  These pairs come in a large set that produce a series of peaks in the diagonal spectrum with decreasing height.  Within a single pair, as the final state in one transition becomes occupied, the peak for that pair is halved.  Then, when the remaining transition becomes blocked out, the peak disappears altogether.  This comes at the same time as a halving of the next peak in the series.  In the off-diagonal spectrum, these pairs cancel each other out.  It is only when an arrow becomes unpaired that a peak appears.

(3) \textit{Flat-band transitions}: These are like the nearest-interband $s=1$ transitions.  Arrows come in equal-energy pairs that share the same final state (for $\mu>0$), forming an isosceles triangle in the snowshoe diagram.  As this state becomes occupied, the peak associated with the pair will wholly disappear.  This happens in a cascading manner along the series as peaks for higher-energy pairs disappear in turn.  In the off-diagonal spectrum, these arrows do not cancel each other out, but instead compete for dominance, depending solely on the relative strength of their overlap function, Eq.~(\ref{eqn:overlap}).

(4) \textit{Asymmetric}: This type was first seen as $\frac{1}{2}\rightarrow\frac{3}{2}$ band transitions in the $s=\frac{3}{2}$ DW.  Instead of creating a series of peaks as in the last two types, these transitions produce individual peaks that will appear and disappear as the chemical potential crosses between initial and final states.  The asymmetry of these arrows means that they do not come in equal-energy pairs.

We have seen that the spectra studied at a fixed chemical potential still have distinct features particular to each DW.  In general, for increasing values of $s$, the spectra become more complicated with several series of peaks, compared to the single series found in the graphene spectra.  In addition, there may be other features that are comparable in part to graphene, but show anomalies.  For example, the longitudinal spectrum for $s=1$ at low doping in Fig.~\ref{fig:sigmaxx1}(a) has a peak at energy $\Omega=2.45\gamma$ (the second peak) which is smaller in height than the peak to its right.  In contrast, the same spectrum for $s=\frac{1}{2}$ has peaks with monotonically decreasing amplitude with increasing energy.  In the $s=1$ low-doping ($\mu>0$) spectrum for right-hand circular polarized light (Fig.~\ref{fig:sigmaminus1}(a)), the peak at $\Omega=2.45\gamma$ is non-existent, while its analogue is present in the $s=\frac{1}{2}$ spectrum (Fig.~\ref{fig:sigmaplus1_2}(a)).  While the features between $s=1$ and $s=\frac{1}{2}$ of the other circular polarization at low-doping are similar, we see a difference between $s=1$ and $s=\frac{1}{2}$ when comparing both polarizations.  In graphene, the series of interband peaks are of equal magnitude in either polarization.  In $s=1$ however, a particular interband peak will have a different amplitude in each polarization.  This is a characteristic of transitions coming from the flat band.

Each DW system creates a unique and predictable signature in its response to optical stimuli.  Absorption experiments, like those already performed on graphene, will thus be useful in classifying any candidate DW system and assigning a specific pseudospin value to it.  In addition, it is necessary to know the features of this response in the design of any optical devices incorporating DW fermions.

\begin{acknowledgements}
This work has been supported by the Natural Science and Engineering Research Council of Canada.
\end{acknowledgements}

\appendix

\section{Derivation of Landau level wavefunctions}\label{sec:Alphas}

The following derivation is very similar to the method employed by Lan \textit{et al}.\cite{Lan11} in finding the Landau energies listed in Table \ref{tab:Landau}.  Here, the technique is used to find the eigenvectors for each of these energy eigenvalues.  Please note that the expressions for the elements $\left\{\alpha^{i}\right\}$ below have not been normalized.  This action must be taken before using said expressions in any subsequent calculations.

Acting the Hamiltonian (\ref{eqn:WHamB}) on the eigenvector (\ref{eqn:ket}) gives the following set of coupled equations, where indices $\lambda$ and $n$ have been supressed and $\epsilon\rightarrow\gamma\epsilon$.
\begin{equation}\label{eqn:App_eigens}
\begin{split}
\epsilon\alpha^{1}=&\rho_{1}\sqrt{n-2s+1}\alpha^{2} \\
\epsilon\alpha^{2}=&\rho_{1}\sqrt{n-2s+1}\alpha^{1}+\rho_{2}\sqrt{n-2s+2}\alpha^{3} \\
&\vdots \\
\epsilon\alpha^{m}=&\rho_{m-1}\sqrt{n-2s+m-1}\alpha^{m-1} \\
&\hphantom{xx} +\rho_{m}\sqrt{n-2s+m}\alpha^{m+1} \\
&\vdots \\
\epsilon\alpha^{2s+1}=&\rho_{2s}\sqrt{n}\alpha^{2s}
\end{split}
\end{equation}
Three distinct cases arise in solving this set of equations.

\subsection{$\epsilon=0$, $2s+1$ odd}

This case deals with the Landau states found within the flat band of an integer-$s$ material.  Plugging $\epsilon=0$ into (\ref{eqn:App_eigens}), one finds that
\begin{equation}\label{eqn:App1_recur1}
\alpha^{m} = \left\{
\begin{array}{c l}
-\frac{\rho_{m+1}\sqrt{n-2s+m+1}}{\rho_{m}\sqrt{n-2s+m}}\alpha^{m+2} & \mathrm{odd}\:m<2s+1 \\
0 & \mathrm{even}\:m
\end{array}\right.
\end{equation}
This is not true however for odd values of the Fock number satisfying $n<2s+1$, where the only solution is the trivial one: $\left\{\alpha^{i}=0\,\,\,\forall\, i\right\}$.  This shows why Landau levels with these indices do not exist in the flat band, as seen in Fig.~\ref{fig:bare}(a) for $s=2$.

By setting a prenormalized value of $\alpha^{2s+1}=1$ subsequent substitutions for the odd elements found in (\ref{eqn:App1_recur1}) above give the following explicit expression.
\begin{equation}\label{eqn:App1_alpha1}
\begin{split}
\alpha^{2m+1}=\left(-1\right)^{s+m}\prod_{r=m}^{s-1}\frac{\rho_{2r+2}\sqrt{n-2s+2r+2}}{\rho_{2r+1}\sqrt{n-2s+2r+1}}&\:,\\
m=0,...,s-1\:&
\end{split}
\end{equation}

\begin{table}
\caption{\label{tab:recursion} Expressions for the first few $A_{m}$ defined in (\ref{eqn:Arecursion}), which are used in building the Landau level eigenvectors.}
\centering
\begin{tabular}{l}
\hline\hline
$A_{1}=1$\\
$A_{2}=\epsilon$\\
$A_{3}=\epsilon^{2}-\rho_{1}^{2}\left(n-2s+1\right)$\\
$A_{4}=\epsilon\left[\epsilon^{2}-\rho_{1}^{2}\left(n-2s+1\right)-\rho_{2}^{2}\left(n-2s+2\right)\right]$\\
$\!\begin{aligned}[t]
A_{5} =& \epsilon^{2}\left[\epsilon^{2} - \rho_{1}^{2}\left(n-2s+1\right) - \rho_{2}^{2}\left(n-2s+2\right) - \rho_{3}^{2}\left(n-2s+3\right)\right] \\
&- \rho_{1}^{2}\rho_{3}^{2}\left(n-2s+1\right)\left(n-2s+3\right)
\end{aligned}$ \\
\hline\hline
\end{tabular}
\end{table}

\subsection{$\epsilon=0$, $2s+1$ even}

This case deals with those zero-energy states shared between two bands in fractional-$s$ materials.  What follows is largely the same as the previous case, except with the role of even and odd elements reversed.  That is, $\alpha^{2m+1}=0$ for $m=0,...,s-\frac{1}{2}$, and the expression for the even-numbered elements is given below, again with $\alpha^{2s+1}=1$.
\begin{equation}\label{eqn:App2_alpha2}
\begin{split}
\alpha^{2m}=\left(-1\right)^{s+m+\frac{1}{2}}\prod_{r=m}^{s-\frac{1}{2}}\frac{\rho_{2r+1}\sqrt{n-2s+2r+1}}{\rho_{2r}\sqrt{n-2s+2r}}&\:,\\
m=1,...,s-\frac{1}{2}\:&
\end{split}
\end{equation}

\subsection{$\epsilon\neq0$}

Finally, this case is for all eigenvectors with nonzero energy.  Going back to (\ref{eqn:App_eigens}), subbing the first line into the second and carrying on in this fashion gives
\begin{equation}
\alpha^{m} = \rho_{m}\sqrt{n-2s+m}\frac{A_{m}}{A_{m+1}}\alpha^{m+1}\:,\:m=0,...,2s
\end{equation}
The factors $A_{m}$ are defined recursively below and listed for $m$ up to $m=5$ in Table \ref{tab:recursion}.
\begin{equation}\label{eqn:Arecursion}
\begin{split}
&A_{1}=1\:,\:A_{2}=\epsilon\:,\\
&A_{m}=\epsilon A_{m-1} - \rho_{m-2}^{2}\left(n-2s+m-2\right)A_{m-2}
\end{split}
\end{equation}
Taking a prenormalized value $\alpha^{2s+1}=A_{2s+1}$ one finds that $\alpha^{2s}=\rho_{2s}\sqrt{n}A_{2s}$ and further substitutions yields the following result.
\begin{equation}
\alpha^{m} = \left(\prod_{r=m}^{2s}\rho_{r}\sqrt{n-2s+r}\right)A_{m}
\end{equation}

\bibliographystyle{apsrev4-1}
\bibliography{bibliography}

\end{document}